\documentclass{aa}  
\usepackage[breaklinks=true,colorlinks,citecolor=blue]{hyperref}
\usepackage{graphicx}
\usepackage{enumerate}
\usepackage{epsfig}
\usepackage{color}
\usepackage{txfonts}
\usepackage{natbib}
\usepackage{multirow}
\usepackage{ulem} 
\definecolor{ao}{rgb}{0.0, 0.5, 0.0}

\newcommand{\kmps}{km~s\ensuremath{^{-1} }\,}
\newcommand{\Msun}{M\ensuremath{_\odot}\,}
\newcommand{\Msunyr}{M\ensuremath{_\odot}~yr\ensuremath{^{-1}}\,}
\newcommand{\Msunpc}{M\ensuremath{_\odot}~pc\ensuremath{^{-2} }\,}
\newcommand{\Msunkpc}{M\ensuremath{_\odot}~yr\ensuremath{^{-1}}~kpc\ensuremath{^{-2} }\,}

\newcommand{\Oo}{\displaystyle}

\newcommand{\bea}	{\begin{array}}
\newcommand{\eea}	{\end{array}}
\newcommand{\beq}	{\begin{equation}}
\newcommand{\eeq}	{\end{equation}}
\newcommand{\ben}	{\begin{eqnarray}}
\newcommand{\een}	{\end{eqnarray}}
\newcommand{\bsq}	{\begin{mathletters}}
\newcommand{\esq}	{\end{mathletters}}

\begin{document}

\title{Bar quenching in gas-rich galaxies}
\titlerunning{Bar quenching in gas-rich galaxies}

\author{S. Khoperskov\inst{1}\thanks{sergey.khoperskov@obspm.fr}, M. Haywood\inst{1}, P. Di Matteo\inst{1}, M. D. Lehnert\inst{2}, F. Combes\inst{3}}
\authorrunning{S. Khoperskov et al.}

\institute{GEPI, Observatoire de Paris, CNRS, 5 place Jules Janssen, 92190 Meudon, France 
\and
Sorbonne Universit\'{e}s, UPMC Univ. Paris 6 et CNRS, UMR 7095, Institut d'Astrophysique de Paris, 98 bis bd Arago, 75014 Paris, France
\and
Observatoire de Paris, LERMA, College de France, CNRS, PSL, Sorbonne
Univ. UPMC, F-75014, Paris, France}

\date{Received ; accepted }
 
\abstract{Galaxy surveys have suggested that rapid and sustained decrease
in the star-formation rate, ``quenching'', in massive disk galaxies
is frequently related to the presence of a bar.  Optical and near-IR
observations reveal that nearly 60\% of disk galaxies in the local
universe are barred, thus it is important to understand the
relationship between bars and star formation in disk galaxies. Recent
observational results imply that the Milky Way quenched about 9--10
Gyr ago, at the transition between the cessation of the growth of the
kinematically hot, old, metal-poor thick disk and the kinematically
colder, younger, and more metal-rich thin disk. Although perhaps
coincidental, the quenching episode could also be related to the formation
of the bar.  Indeed the transfer of energy from the large-scale shear
induced by the bar to increasing turbulent energy could stabilize the
gaseous disk against wide-spread star formation and quench the galaxy.

To explore the relation between bar formation and star formation in
gas rich galaxies quantitatively, we simulated gas-rich disk isolated
galaxies.  Our simulations include prescriptions for star formation,
stellar feedback, and for regulating the multi-phase interstellar
medium. We find that the action of stellar bar efficiently quenches
star formation, reducing the star-formation rate by a factor of $10$ in
less than $1$~Gyr. Analytical and self-consistent galaxy simulations
with bars suggest that the action of the stellar bar increases the
gas random motions within the co-rotation radius of the bar. Indeed,
we detect an increase in the gas velocity dispersion up to $20-35$~\kmps
at the end of the bar formation phase. The star-formation efficiency
decreases rapidly, and in all of our models, the bar quenches the star
formation in the galaxy. The star-formation efficiency is much lower
in simulated barred compared to unbarred galaxies and more rapid bar
formation implies more rapid quenching.}


\maketitle

\section{Introduction}\label{sec::intro}

Understanding the formation and evolution of disk galaxies requires,
at a minimum, a detailed knowledge of the star-formation history of the
ensemble of galaxies. When classified by the relative rates of star
formation, galaxies divide into two separate sequences, a red and a
blue sequence consisting of ellipticals and lenticulars and spirals and
irregulars, respectively \citep{2001AJ....122.1861S, 2004ApJ...608..752B}.
Several mechanisms have been proposed to explain the observed transition
from blue active star-forming galaxies to red galaxies.  These range
from invoking the energy output of active galactic nuclei, major mergers
initiating starbursts which exhaust the gas, to accreting gas being
heated to such a high temperatures that it no longer cools significantly
in a Hubble time, to environment effects such a tidal and ram-pressure
stripping, to simple gas exhaustion, or the action of bars in disk
galaxies. Several of these mechanisms may operate in coordination or
simultaneously, perhaps even within the same galaxy.

In our current level of understanding of galaxy evolution, active
galactic nuclei~(AGN) feedback can suppress star-formation
activity \citep{1998A&A...331L...1S, 2005Natur.433..604D,
2012ARA&A..50..455F,2017NatAs...1E.165H}.  A number of galaxy
formation models \citep{2005ApJ...620L..79S, 2006MNRAS.370..645B,
2008MNRAS.391..481S, 2010MNRAS.406.2325O, 2017MNRAS.465..547P}, isolated
galaxy simulations \citep{2014MNRAS.440.2333D,2014MNRAS.441.1615G} and
semi-analytic models \citep{2006MNRAS.365...11C,2008MNRAS.391..481S}
have suggested that AGNs are able to suppress star formation rapidly
either by removing gas from the galaxy~\citep[see e.g.,][and
references therein]{2007MNRAS.380..877S,2013MNRAS.433.3297D},
or by injecting turbulence which stabilizes the gas against
fragmentation and becoming self-gravitation on any scale
\citep{2015A&A...574A..32G,2016ApJ...826...29L}. Various types of
observations confirm that star formation can be suppressed by a factor
of 3-50 \citep[e.g.,][]{2015MNRAS.453L..83M, 2015MNRAS.452.1841S}

It has long been known that galaxy properties are closely related to
the local environmental density~\citep{2004MNRAS.353..713K}. In dense
environments, star formation can be easily quenched by gas removal due
to ram-pressure or tidal interactions~\citep{2011ApJ...742..125G,
2010ApJ...714.1779V,2010AJ....140.1254W,2011MNRAS.415.1797C}.
\cite{2017MNRAS.464..121S}, for example, found that the specific
star-formation rate drops by a factor of $4$ in dense environments.

\cite{2009ApJ...707..250M} analyzed the stability of disks proposing
``morphological'' quenching whereby the formation of a stellar spheroid
stabilizes the disk against gravitational instability and formation of
star-forming clumps. \cite{2013MNRAS.432.1914M} subsequently confirmed
a possible role of disk stabilization in a study of local galaxies.
A strong advantage of this mechanism is that it can explain the
quenching without removing or depleting the gas.  However it is not
clear if this mechanism is efficient in high redshift galaxies where
the fraction of gas in the disk is high, 40-50\%.  While morphological
quenching may be effective in suppressing star formation in early-type
galaxies, it is not the mechanism responsible for quenching observed in
late-type disks, since they lack any significant spheroid~\citep[see,
for example][]{2007MNRAS.381..401L}.

Bars are a common feature in the inner regions of nearby
disk galaxies. Roughly $60\%$ of disk galaxies in the local
universe have  stellar bars~\citep{2000AJ....119..536E,
2007ApJ...657..790M, 2000ApJ...529...93K, 2007ApJ...659.1176M,
2007AJ....134.2286H}.  Galactic bars very likely play an important
role in both secular evolution of disk galaxies \citep[][and
references therein]{1981A&A....96..164C, 1990A&A...233...82C,
2004ApJ...604L..93D, 2005MNRAS.363..496A, 2013seg..book..305A,
2015A&A...577A...1D, 2015A&A...584A..90G}, and in a dynamical
(re-)distribution of gas \citep{1985A&A...150..327C, 1992MNRAS.259..345A,
2000ASPC..221..243A, 2007A&A...472...63R,2007ApJ...666..189B}
and metals in the galactic disks \citep{2004MNRAS.355.1251L,
2013MNRAS.431.2560M,2013A&A...553A.102D,2016MNRAS.460.3784S}. The
dynamics of gas in a bar potential is complex, depending on the
local environment of the ISM and how the bar evolves \citep[see
e.g.,][]{1992MNRAS.259..345A, 1995ApJ...449..508P, 2000ApJ...528..677E,
2001PASJ...53.1163W, 2002MNRAS.329..502M, 2004ApJ...600..595R,
2016MNRAS.462L..41F}. The formation and dynamics of bars is potentially an
important mechanism for regulating the evolution of the star-formation
rate in disk galaxies. While studies have highlighted a possible
link between the star formation and the existence of bars in disk
galaxies \citep[see e.g.,][]{2004ApJ...607..103L, 2005ApJ...630..837J,
2010MNRAS.405..783M, 2011MNRAS.416.2182E}, by studying the central
regions of four strongly barred galaxies, \cite{2016MNRAS.457..917J}
showed that the recent star formation appears to have been suppressed by
the bars. By reconstructing the star-formation history of main-sequence
galaxies from $z=3$ to the present epoch, \cite{2015A&A...580A.116G} found
that star-forming galaxies quenched above a mass threshold which increases
with increasing redshift. They also noted that in the local Universe there
is a sharp increase in the fraction of visually classified strong bars
as a function of mass, suggesting that strong bars may be responsible
for the quenching observed at high redshifts and bars ability to quench
star formation depends on galaxy mass. In such a scenario, the bar sweeps
most of the gas into the galactic center, where it is then converted into
stars. The vigorousness of the resulting bar-induced starburst depends
on the mass of the galaxy, with massive barred galaxies converting all
the gas funnelled to their centers into stars \citep{2016MNRAS.463.1074C}.

An analysis of the ErisBH simulation demonstrated that the formation of a
bar in a galaxy with a quiet merger history can quench its star formation
on kiloparsec scales~\citep{2017MNRAS.465.3729S}. They showed that gas
can be removed rapidly by the bar in the inner region, preventing any
further strong star formation. \cite{2013ApJ...779..162C} also discussed
the possibility that bars are able to quench star formation. Through
an analysis of Galaxy Zoo~2 dataset they claim that secular evolution
is able to build high enough central densities to act as a quenching
mechanism. Complex cases may occur though. Observational
results for the galaxy NGC~4371 suggests that, although barred,
the quenching of its star formation is very likely an environmental
effect~\citep{2015A&A...584A..90G}. Thus, possible complications due to
environmental effects make it even harder to extract a general statement
on the ways star formation and quenching proceed in barred galaxies.

Since a large fraction of galaxies have a bar at the present epoch and
many galaxies~\citep[including the Milky Way, see][]{2016A&A...589A..66H}
passed through a stage of quenching, we want to study the possible
relation between the bar formation and quenching phase for Milky
Way-type galaxies. This link is intriguing, in particular in the Milky
Way, where it has been established that a drastic (about a factor
$10$) and rapid (in about $1$~Gyr) decrease in the star-formation rate
occurred approximatively $9-10$~Gyr ago, at the transition between the
formation of the thick and thin disks of our Galaxy. For galaxies with
masses similar to the Milky Way, this epoch corresponds also to the
time when a substantial increase in the fraction of barred galaxies
is observed~\citep{2008ApJ...675.1141S,2012ApJ...758..136S}. It is
thus important to understand and quantify if there exists a causal
link between these two phenomena.  \citet[][]{2016A&A...589A..66H},
for example, proposed that in the Milky Way the action of a stellar
bar could have increased the gas turbulence, stabilizing the disk
against star formation, quenching its star formation. In this scenario,
the suppression of the star formation is not associated to a substantial
consumption of the available gas, as in other proposed mechanisms. Indeed,
the gas would still be present in the disk at those epochs, but its high
turbulence would have significantly limited the star-formation efficiency.
Not depleting the gas reservoir is the important element, which, at
least for the quenching episode in the Milky Way, has to be the
case. There is a continuity in the elemental abundances between the
stellar populations that formed before and after the quenching episode,
which implies that the gas reservoir was not substantially replenished
after the quenching episode~\citep[][in prep.]{2016A&A...589A..66H,2017Haywood}.

The present study aims to address the link between bar formation
and star-formation rate in gas-rich galaxies. To accomplish this,
we analyze  a suite of simulated galaxies with a range of bar
parameters (strength, formation time-scales) and different star
formation prescriptions.  The paper is structured as follows. We
describe our simulation code in Section~\ref{sec::code} and our model
simulations in Section~\ref{sec::model}. Results are presented in
Section~\ref{sec::results} where we compare star-formation histories,
star-formation efficiency, gas velocity dispersion in barred and
unbarred galaxies. We also analyse a set of simulations with various bar
parameters and describe our self-consistent ``live'' bar simulations. In
Section~\ref{sec::discuss} we discuss our results in the context of the
evolution of galaxies and the Milky Way. We summarize the key results
of our investigation in Section~\ref{sec::concl}.

\section{Simulations}
\subsection{Models}\label{sec::code}

We study star formation in barred galaxies via $N$-body/hydrodynamical simulations of the stellar-gaseous disk of Milky Way-like galaxies embedded in a dark matter halo potential. We use the code based on the TVD MUSCL scheme~(Total Variation Diminishing Multi Upstream Scheme for Conservation Laws) for gas dynamics and TreeCode for gravity calculation \citep{2014JPhCS.510a2011K}. We incorporate a multi-phase model of the ISM and star-formation prescription \citep{2013MNRAS.428.2311K, 2016MNRAS.455.1782K}.  We adopted gas cooling and heating rates from~\cite[][see Appendix B]{2013MNRAS.428.2311K}. This model produces a multi-phase ISM with $T\approx 100$~K for cold phase, $T\approx 10^4$~K for warm medium and $T\approx10^6$~K for hot medium. In these simulations we adopted a gas radiative cooling for metallicity $0.5 Z_\odot$. Star formation from the cold phase gas is based on three criteria \textit{(i)} Jeans mass criterion ~($M_{\rm gas}>M_{\rm Jeans}$); \textit{(ii)} temperature threshold ($T<100$~K); and \textit{(iii)} converging flow $\nabla \cdot v<0$. We create a stellar particle in a gaseous cell when hydrodynamical variables satisfy all these star formation criteria. Note, in particular, that the adoption of a convergent flow criterion in star formation prescription is widely accepted in the literature~\cite[see for instance][]{2011MNRAS.417.1318D, 2014MNRAS.445.3352C, 2015MNRAS.451.4223M, 2016ApJ...821...90A}. These works also use converging flow criteria for star formation in isolated galaxy simulations and they have a spatial resolution in the range of~$40-70$~pc.

In our simulations the star particles are assumed to represent a stellar cluster whose mass distribution follows a Salpeter initial mass function
\citep{1955ApJ...121..161S}. Mass loss by stellar populations as well as SNe energy ejection at each time step are calculated according to
stellar evolution code STARBURST99 \citep{1999ApJS..123....3L}. We do not consider central black hole formation/evolution in our simulations.

In the paper we consider two types of models: models with a rigid
disk and dark matter halo potential and self-consistent simulations.
For models with a rigid potential, the total gravitational potential
$\Psi_{\rm tot}$ is given by:
\begin{equation}
\Oo \Psi_{\rm tot}({\bf r}, t) = \Psi_{\rm h}({\bf r}) + \Psi_{\rm d}({\bf r}) \times (1 + \zeta(t) \Psi_{\rm b}({\bf r}) )\,,    
\end{equation}
where ${\bf r}$ is the space coordinate vector, $t$ is the time variable,
$\Psi_{\rm h}$ is the dark matter halo in pseudo-isothermal sphere
model \citep{1995ApJ...447L..25B}, $\Psi_{\rm d}$ is the stellar disk
potential, defined through the Bessel functions~\citep[see e.g. Chapter
2.6.2 in][]{2008gady.book.....B}, $\Psi_{\rm b}$ is the bar-perturbation
and $\zeta(t)$ is the bar amplitude. These functions are adopted
according to~\cite{2000AJ....119..800D}:
\beq
	\Psi_{\rm b} = \cos\left(2[\varphi-\Omega_b t]\right)\,
		\times\left\{ \bea{l@{$\quad$}l}
		-(r_b/r)^3	& {\rm if}\;r\ge r_b, \\[2ex]
		(r/r_b)^3-2	& {\rm if}\;r\le r_b, \eea \right.
\eeq
The amplitude of non-axisymmetric perturbation $\zeta$ is switched on smoothly. It is zero before $t=0$,
grows with time at $0<t<h$ as
\beq \label{ampl}
	\zeta(t) = \varepsilon_b \left({3\over16} \xi(t)^5 - {5\over8} \xi(t)^3  + {15\over16} \xi(t) + {1\over2}\right), \quad \xi(t) \equiv 2 {t\over h} - 1,
\eeq
and stays constant at $\zeta(t)=\varepsilon_b$ after a time, $h$. 
The bar is assumed to rotate rigidly about the galactic center with a fixed
pattern speed ($\Omega_b \approx 52$~\kmps)  and a co-rotation radius close to
the end of bar, which is in agreement with previous studies~\citep[see
e.g.][]{1980A&A....81..198C,1993A&A...271..391C}.

In most of the models run
with a rigid halo and stellar disk potential, we also introduce a bar
perturbation, whose amplitude increases according to the function
$\zeta(t)$ defined above. To quantify the effect of the bar on the star-formation history of the simulated disk galaxy, we also run some
rigid models, without including any asymmetric perturbation (unbarred
models). In all these models (barred and unbarred), the potential
is initially axisymmetric, and so is the gas density distribution.
For generating the vertical and radial equilibrium for the gaseous disk,
we used the technique described in \cite{2012MNRAS.427.1983K}. 

For the self-consistent simulations we consider models with initial
stellar and gaseous disk embedded into a live dark matter halo
potential. The density profiles of the dark halo is a Plummer
sphere and initial stellar disk follows a Miyamoto-Nagai profile
\citep{1975PASJ...27..533M}. The galaxy is composed of $10^6$ particles
of ``dark matter'', and $10^6$ ``stellar particles'' for the initial
disk. Equilibrium distribution functions for these components are found by
using the method described in \cite{2009MNRAS.392..904R}. For the gaseous
disk component (for both types of models) we adopted a computational
domain $30 \times 30 \times 8$~kpc with a uniform grid and a spatial
resolution of $50$~pc. All parameters~(masses, spatial scales and etc.) of
our models are listed in Table~\ref{tab::table_ini}.  We set the opening
angle of the TreeCode to $\theta = 0.5$ in all simulations. Since we take
into account star formation, the final number of stellar particles depends
on the star-formation rate and it varies in the range $(2-4)\times10^6$
after $2$~Gyr of evolution.

\subsection{Parameter regimes}\label{sec::model}

We performed a set of isolated galaxy simulations, evolving them for
$2$~Gyr since we are only interested in modeling bar formation and its
impact on the star-formation history, rather than their long term,
secular evolution. We simulated a total of $12$~models: $2$~rigid, axisymmetric
models; $8$~rigid asymmetric models with a bar whose strength can vary
from $\varepsilon_{\rm b}=0.1$ to $\varepsilon_{\rm b}=1$, and whose
formation timescale has been varied from $h=0.2$~Gyr to $h=1$~Gyr;
$2$~self-consistent simulations, with a live halo and live initial stellar
disk. All model parameters are given in Table~\ref{tab::table_ini}. For
all models, we used the star-formation prescription described
previously, except for the three models marked with minus in
Table~\ref{tab::table_ini}. For these three simulations, we used a
star-formation prescription without the converging flow criterion. As
we will discuss subsequently, not including this criterion allows us to
clarify the role of random gas motions on the calculated star-formation
history~(SFH). We start our analysis with two fiducial models (barred RB and unbarred RA) and then the effect of varying parameters is studied by using further 8 models with rigid bar potential. Finally we discuss our ``self-consistent'' simulations.

\begin{table*}
\begin{center}
\caption{Initial parameters: $M_h$ is the halo mass, $a_h$ is the halo
scale length;  $M_d$, $R_d$, $z_d$ are the initial stellar disk mass,
radial scale length and vertical scale height respectively; $M_{\rm gas}$,
$R_{\rm gas}$, $z_{\rm gas}$ are the initial stellar disk mass, radial
scale length and vertical scale hight respectively; $r_{\rm b}$ is the
bar size, $\varepsilon_{\rm b}$ is the bar potential amplitude, $h$ is
the bar growth timescale in models with a rigid bar potential. Models
marked by minus include no converging flow criterion in the
star-formation prescription. First section of the table shows the model
with no bar imposed in the disk (axisymmetric potential); the second
section presents models with a rigid potential of bar; and the last section
shows the parameters of the self-consistent simulations that have
live halos and an initial stellar disk. For simulations with a
rigid potential, we adopted an isothermal dark matter~(iso) distribution
and exponential~(exp) stellar disk; for self-consistent
simulations we used Plummer~(plum) sphere for dark matter halo and
Miyamoto-Nagai~(M-N) initial stellar disk. \label{tab::table_ini}}

\begin{tabular}{lccccccccccccccccccc}
\\
\multicolumn{1}{c}{Model} &\multicolumn{3}{c}{Halo}&\multicolumn{4}{c}{Initial stellar disk}&\multicolumn{3}{c}{Gas}&\multicolumn{3}{c}{Bar parameters}\\
\hline
 		&type & $M_h$ & $a_h$ & type & $M_d$& $R_d$ & $z_d$ & $M_{\rm gas}$ & $R_{\rm gas}$ & $z_{\rm gas}$ &  $h$ & $r_{\rm b}$ & $\varepsilon_{\rm b}$\\
 		&  & $10^{10} M_\odot$ & kpc &   & $10^{10} M_\odot$ & kpc & kpc & $10^{10} M_\odot$ & kpc & kpc &  Gyr & kpc & \\
\hline		
RA	&rigid / iso &5.8 &5 &rigid / exp&2 & 3 & 0.3 & 2 & 6 & 0.2 & - & - & - \\
RA-	&rigid / iso &5.8 &5 &rigid / exp&2 & 3 & 0.3 & 2 & 6 & 0.2 & - & - & -\\
\hline
RB	&rigid / iso &5.8 &5&rigid / exp&2 & 3 & 0.3 & 2 & 6 & 0.2 & 0.2  & 5 & 0.1\\
RBm02	&rigid / iso &5.8 &5&rigid / exp&2 & 3 & 0.3 & 0.2 & 6 & 0.2 & 0.2  & 5 & 0.1\\
RB-	&rigid / iso &5.8 &5&rigid / exp&2 & 3 & 0.3 & 2 & 6 & 0.2 & 0.2  & 5 & 0.1\\
RBh05	&rigid / iso &5.8 &5&rigid / exp&2 & 3 & 0.3 & 2 & 6 & 0.2 & 0.5  & 5 & 0.1\\
RBh1	&rigid / iso &5.8 &5&rigid / exp&2 & 3 & 0.3 & 2 & 6 & 0.2 & 1  & 5 & 0.1\\
RBe02	&rigid / iso &5.8 &5&rigid / exp&2 & 3 & 0.3 & 2 & 6 & 0.2 & 0.5  & 5 & 0.2 \\
RBe05 	& rigid / iso & 5.8 &  5 & rigid / exp &2 & 3 & 0.3 & 2 & 6 & 0.2&  0.5  & 5 & 0.5   \\
RBe1 	& rigid / iso & 5.8 &  5 & rigid / exp &2 & 3 & 0.3 & 2 & 6 & 0.2&  0.5  & 5 & 1.0   \\
\hline
LB  & live / plum & 10 &  10 & live / M-N &2 & 3 & 0.3 & 2 & 6 & 0.2 & -  & - & -   \\
LB- & live / plum & 10 &  10 & live / M-N &2 & 3 & 0.3 & 2 & 6 & 0.2 & - & -  & -   \\% no converging flow  \\
\end{tabular}
\end{center}
\end{table*} 

\section{Results of simulations}\label{sec::results}
\subsection{Barred versus unbarred models : gas distribution and star forming regions}

%%%%%%%%%%%%%%%%%%%%%%%%%%%%%%%%%%%%%%%%%%%%%%%%%%%%%%%%%%%%%%

\begin{figure*}
\includegraphics[width=1\hsize]{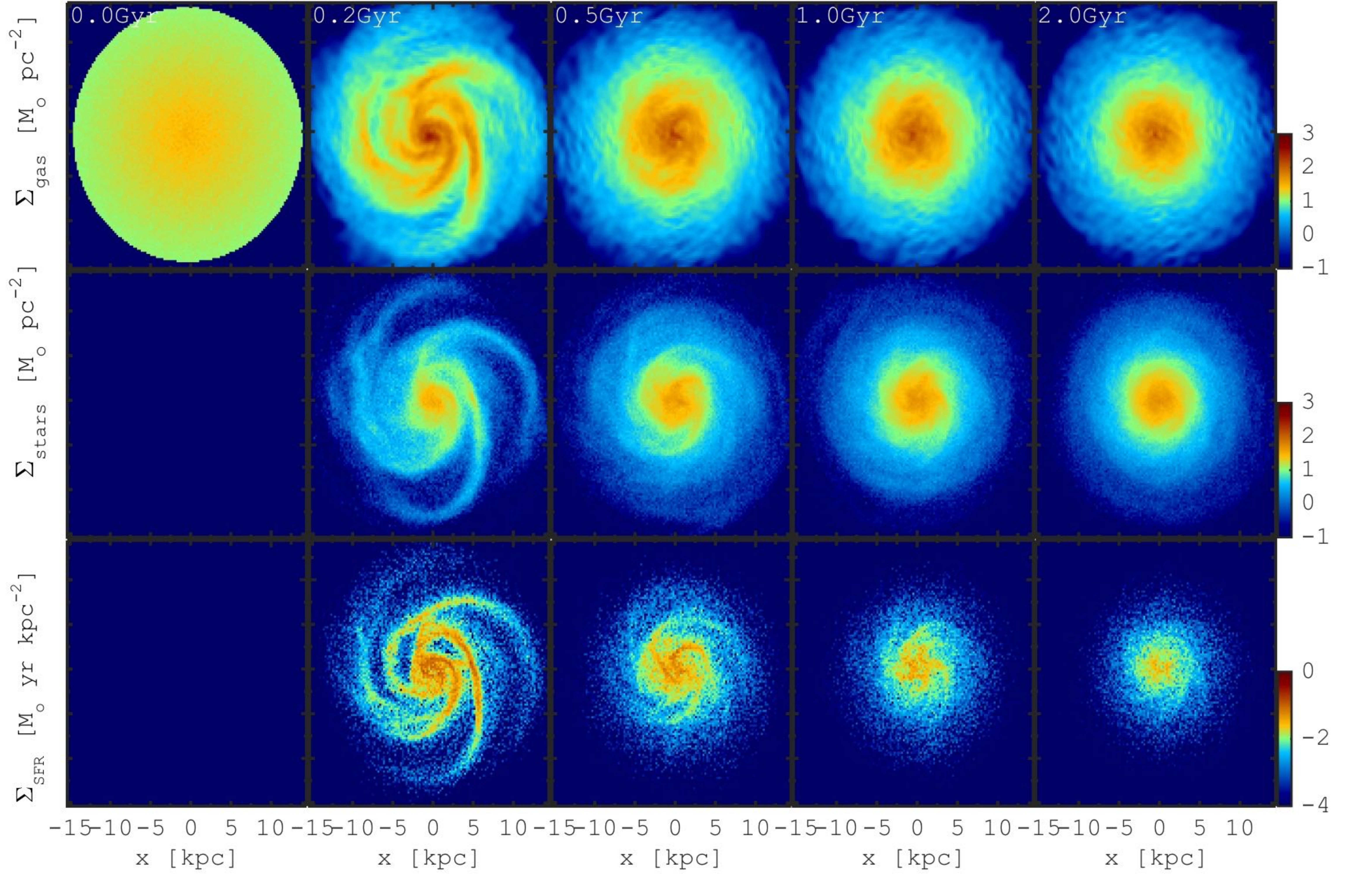}
\caption{The evolution of a gas surface density \textit{(top row)},
surface density of young stars \textit{(middle row)} and a mean star
formation rate over the last $5$~Myr \textit{(bottom row)} in model without
a bar potential. Rotation is anti-clockwise in each panel. The each box
size of the face-on maps is $30\times30$~kpc$^2$.}
\label{fig::evolution_no_bar}
\end{figure*}

%%%%%%%%%%%%%%%%%%%%%%%%%%%%%%%%%%%%%%%%%%%%%%%%%%%%%%%%%%%%%%

Figure~\ref{fig::evolution_no_bar} shows the distribution of stars,
gas and star-formation surface density at four different times, for
the rigid, axisymmetric model. At a simulation time of $t = 0.2$~Gyr,
flocculent spiral arms are clearly visible in both components. These
structures have high cold gas fractions and are sites of enhanced
star formation. Stars form throughout the whole disk, with star-formation rate
surface-density following the gas distribution and recently formed stars trace
the gas structures well. After this initial phase, lasting for less than
$1$~Gyr, the large scale structures dissipate rapidly and the models then
follow a quasi-steady evolution of the disk. Flocculent spiral structures
are short lived because the disk tends to be heated up by the
action of the spiral waves \citep[see e.g.,][]{1991A&A...250..316T,
1985ApJ...292...79C, 2011MNRAS.410.1637S} which in turn stabilizes the
disk against non-axisymmetric perturbations.

%%%%%%%%%%%%%%%%%%%%%%%%%%%%%%%%%%%%%%%%%%%%%%%%%%%%%%%%%%%%%%
\begin{figure*}
\includegraphics[width=1\hsize]{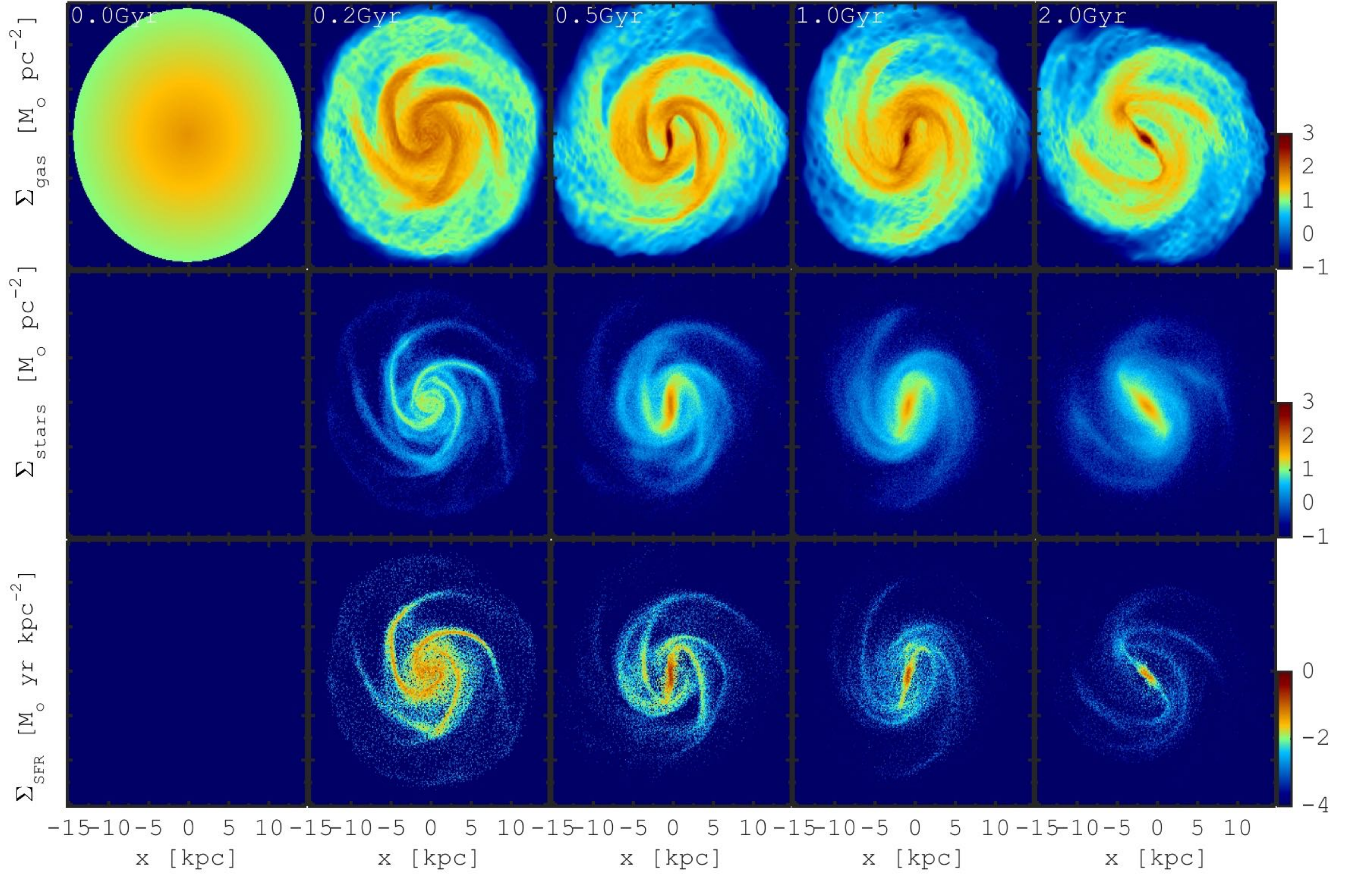}
\caption{As in Fig.~\ref{fig::evolution_no_bar}, but for the
model with a rigid bar potential ($\varepsilon_b=0.1$, $r_b = 5$~kpc,
$h=0.2$~Gyr).}
\label{fig::evolution_rigid_bar}
\end{figure*}
%%%%%%%%%%%%%%%%%%%%%%%%%%%%%%%%%%%%%%%%%%%%%%%%%%%%%%%%%%%%%%

The typical evolution of a rigid model with the bar potential imposed
in the disk is shown in Fig.~\ref{fig::evolution_rigid_bar}. The barred
model shown is the model RB (Table~\ref{tab::table_ini}), having $h=
0.2$, $\varepsilon_{\rm b}=0.1$. There is a strong redistribution of the
gas at initial stages of evolution~($t<0.2-0.5$~Gyr), coinciding with
the formation of the bar, and after $0.5$~Gyr a bar pattern is clearly
visible both in the gaseous and young stellar component. Relative to
the size of the bar, an extended outer spiral structure also arises
in the disk. The bar is able to influence the spiral pattern outside
of its co-rotation. At later times, star formation mostly occurs
over a small fraction of the disk, mainly along the bar major axis.
Recently formed stars also trace similar patterns and form both along
the bar and the trailing spiral structure which is still visible after
$2$~Gyr of evolution. The spiral structure appears to be much weaker in
the underlying stellar distribution. There is also a strong, $\approx
10^9$~\Msun, mass concentration of gas in the very center ($<1$~kpc)
because the stellar bar potential drives a gas flow inside the
inner Lindblad resonance~\citep[see e.g.][]{1992MNRAS.259..345A}. At the
leading side of the bar there are shocks which at least observationally,
coincide with the dust lanes, and are extremely narrow. In spite of the
fact that we are using an analytic gravitational potential for the bar in
this model, a variety of structural properties of the galaxy are evident
from the snapshots, including ISM morphology, qualitatively similar to
those observed in actual barred galaxies.

%%%%%%%%%%%%%%%%%%%%%%%%%%%%%%%%%%%%%%%%%%%%%%%%%%%%%%%%%%%%%%
\begin{figure*}
\includegraphics[width=0.5\hsize]{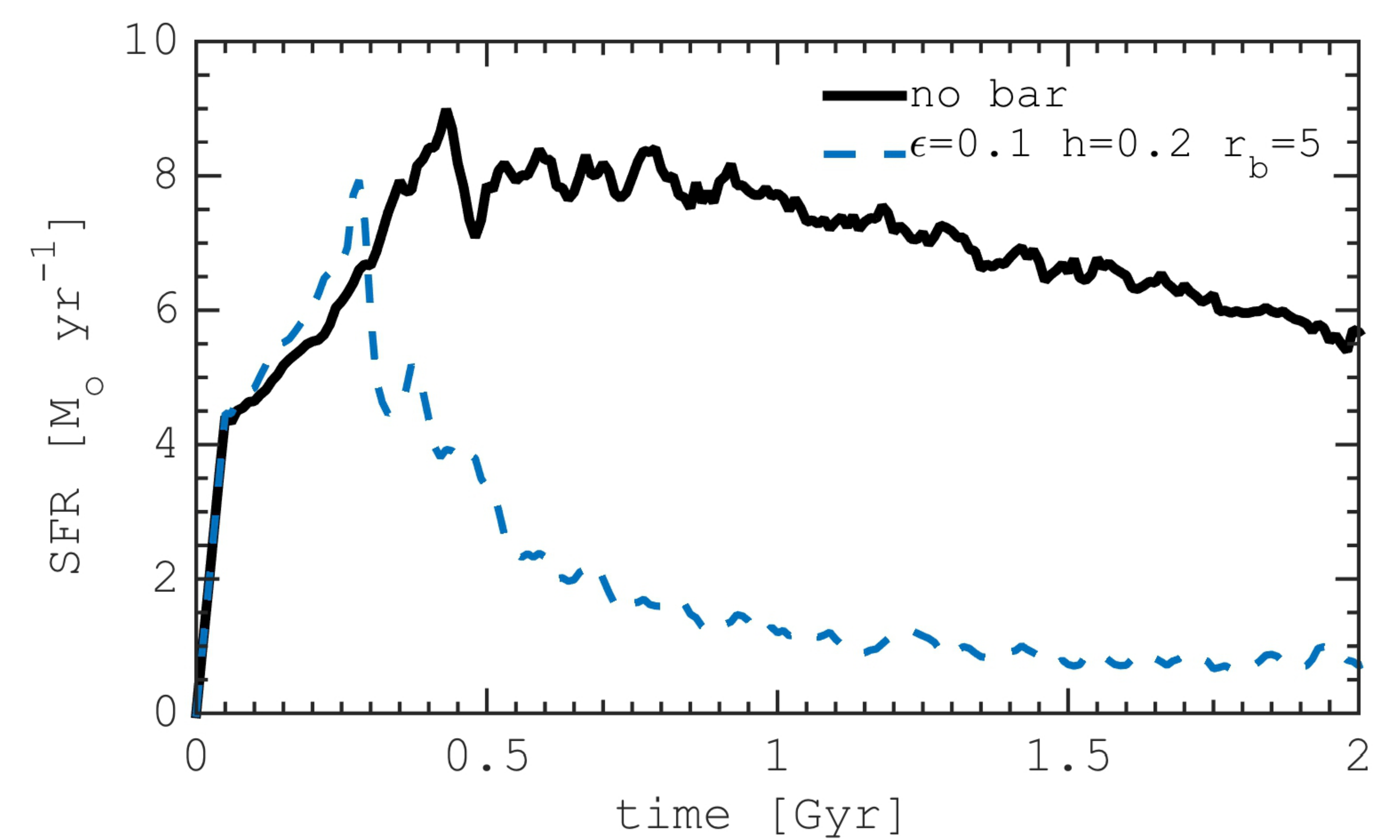}
\includegraphics[width=0.5\hsize]{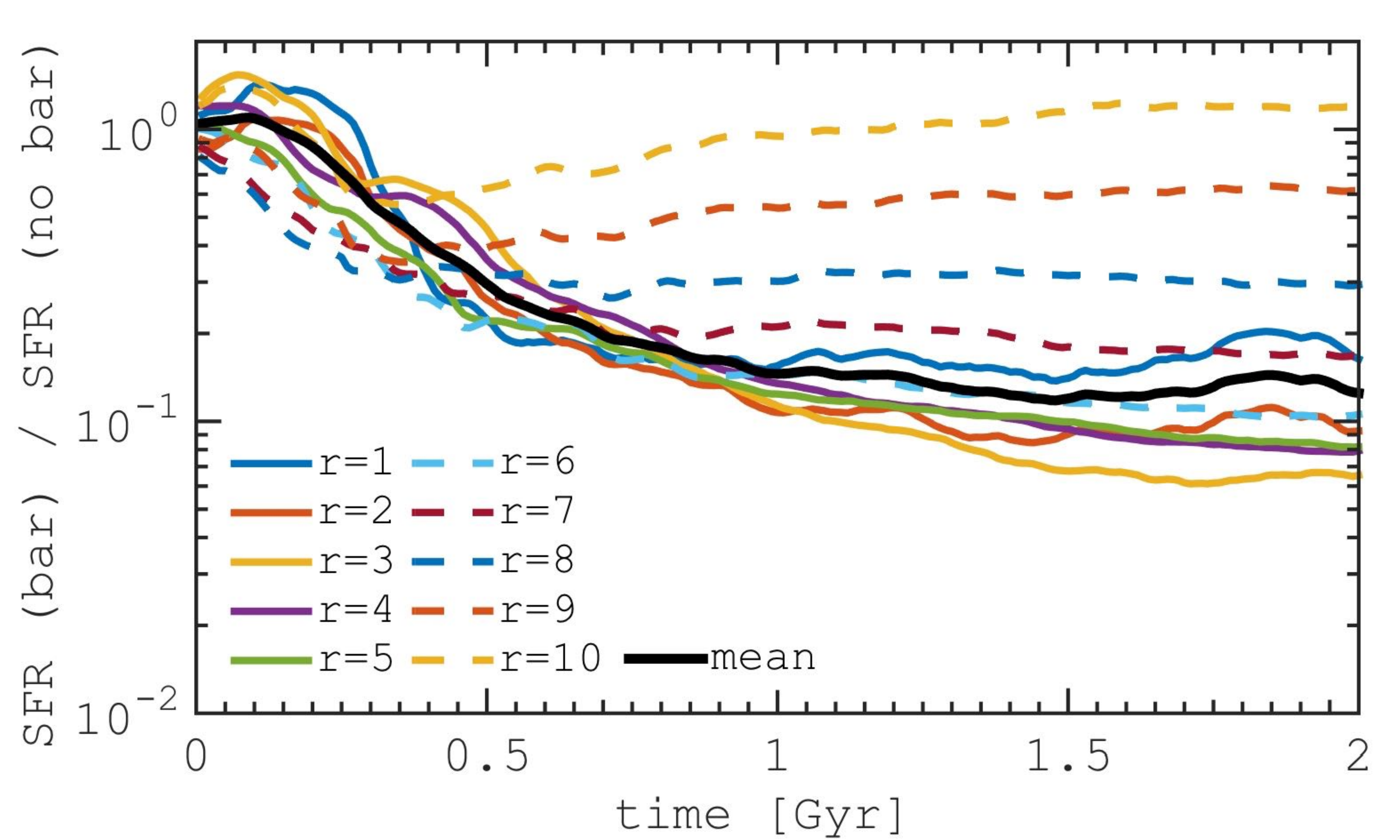}
\caption{\textit{Left:} Comparison of the global star-formation rate
in a model with a bar potential imposed in the disk~(dashed line) and
in a model with axisymmetric potential~(black line). Bar parameters
are given in the legend. \textit{Right:} star-formation rate ratio for
these models. Different lines correspond to the ratio at different radii.
Solid lines represent the results for radii over the  range $1-5$~kpc and
dashed lines the radii over the range $6-10$~kpc. Each line corresponds
to the SFR calculated in a ring of $1$~kpc width. Black solid line
represents the mean star-formation rate ratio over the whole disk.}
\label{fig::sfr_global} \end{figure*}
%%%%%%%%%%%%%%%%%%%%%%%%%%%%%%%%%%%%%%%%%%%%%%%%%%%%%%%%%%%%%%

The star-formation histories of the simulated barred and
unbarred galaxies summarize the efficiency of gas consumption
for models galaxies with different (imposed or live) disk
structures. Figure~\ref{fig::sfr_global}~(left) shows the global
star-formation rate, i.e., that over the whole disk, in the two models
with and without a bar as a function of time. Once the simulation starts,
both models show that the star-formation rate increases rapidly reaching
up to $4$~\Msunyr during the first $0.1$~Gyr. Then the feedback from
stars reduces the rate at which star formation increases. However, the
gas reservoir is still very large and the star-formation rate continues
to increase up to $8$~\Msunyr between $0.3-0.5$~Gyr.  After $0.5$~Gyr,
the star-formation rate of the unbarred galaxy varies very slowly. It
only decreases by a factor of $1.25$ over $2$~Gyr after its initial
peak, remaining at the level of $6-8$~\Msunyr. This model has steady
conversion of gas into stars without the formation of any prominent
spatial structures. It is clear that the model does not possess any
rapidly growing instabilities.

When a bar is present in a model with the same structure, the model
shows a rapid decrease in the star-formation rate immediately after
the bar amplitude reaches its maximum strength. Such decrease starts
at $t=0.2-0.3$~Gyr for the reference model~(RB). Star formation in
the model decreases rapidly to $\approx$1~\Msunyr within $1$~Gyr. After
$1$~Gyr, the star-formation rate is roughly constant. The gas in the very
inner regions is  compressed by the bar potential and depleted by star
formation.  The star formation in the outer disk is sufficiently low
that the distribution and intensity of the star formation there
disk changes very little throughout the simulation. The bar has little
influence on the star-formation rate in the outer disk.

Figure~\ref{fig::sfr_global} shows the ratio between the star-formation
rates at different radii in the two simulations with and without bars. In
the case of the simulation with a bar, the star-formation rate decreases
only in the inner region.  Inside $5$~kpc, where the bar action is the
dominant driver of dynamical evolution, star formation drops down by
a factor of $5-30$. In the outer disk the star formation in the barred
galaxy is more similar to that of the unbarred case -- the ratio between
the barred and unbarred simulated galaxies is close to unity. However,
the SFR at larger radii remains relatively low and the global quenching
observed in the model is largely driven by the star-formation rate
evolution in the central regions of the disk. When we average over the
entire extent of the disk, the star-formation rate is a factor of $10$~lower than that observed in the axisymmetric case.

\subsection{Barred versus unbarred models: evolution with time of the
star-formation rate, gas mass and star-formation efficiency}

%%%%%%%%%%%%%%%%%%%%%%%%%%%%%%%%%%%%%%%%%%%%%%%%%%%%%%%%%%%%%%
\begin{figure*}
\includegraphics[width=0.49\hsize]{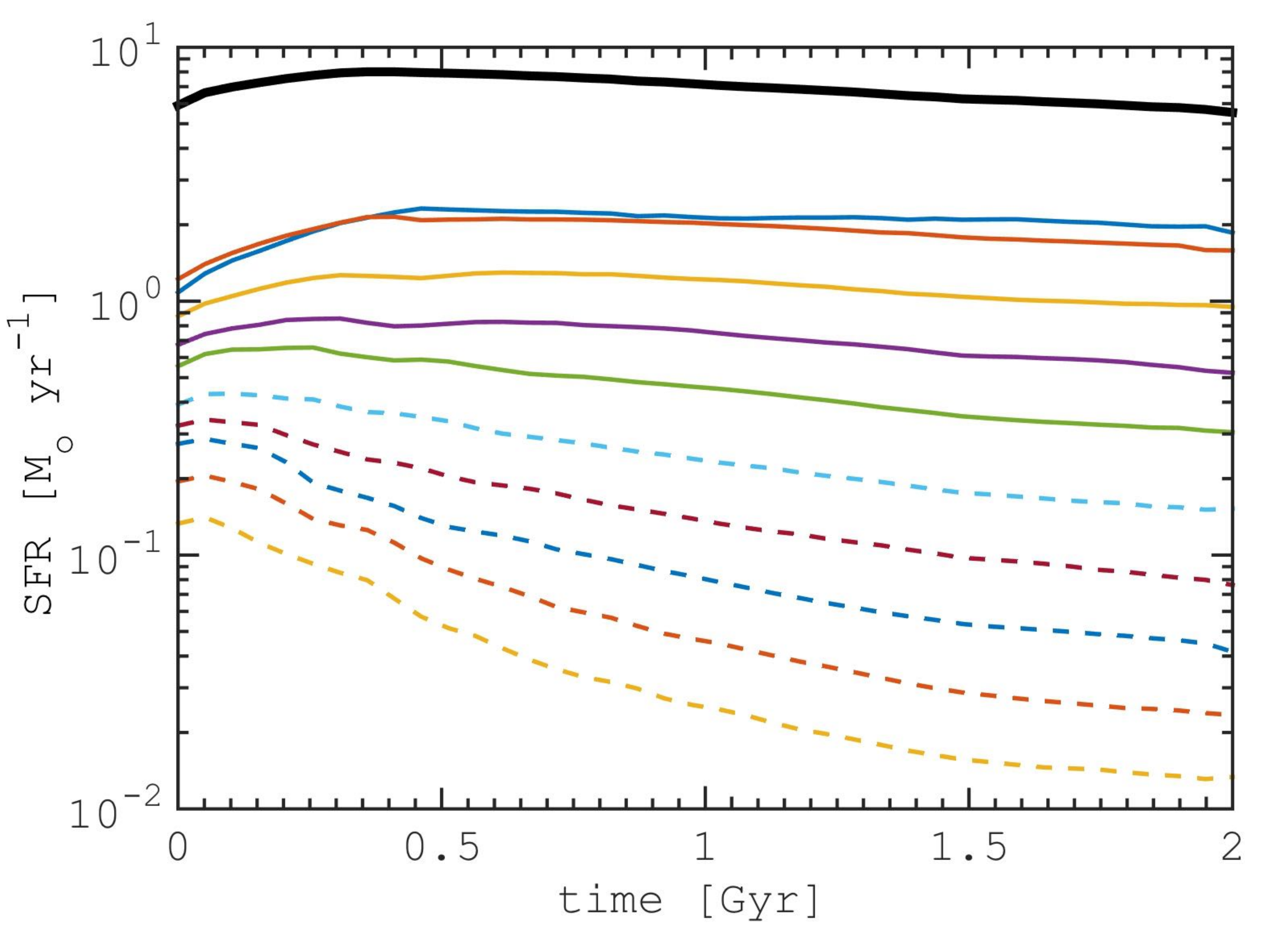}\hfill\includegraphics[width=0.49\hsize]{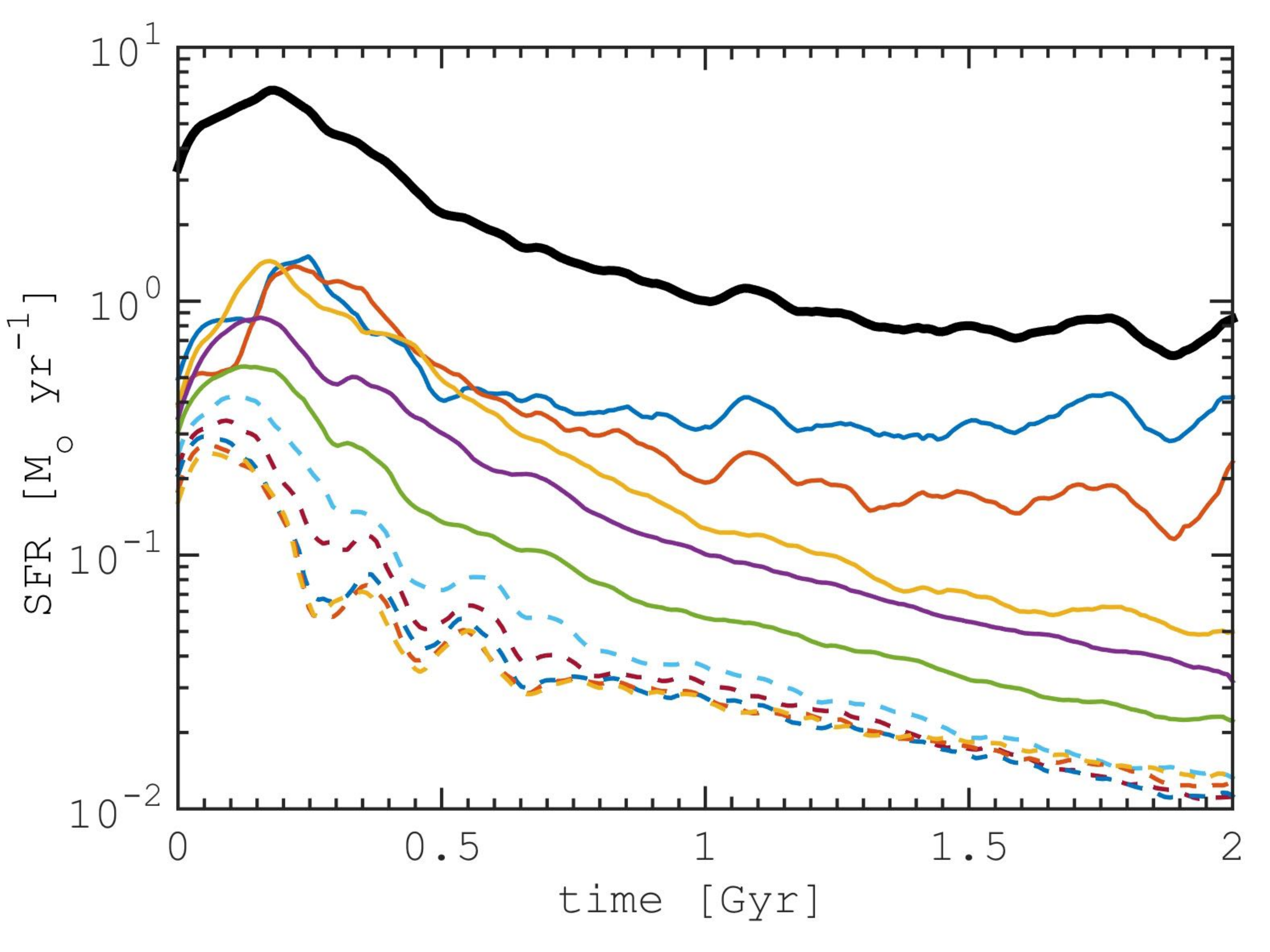}
\caption{Star-formation rate in different models: one with no
bar \textit{(left)}, one with a bar \textit{(right)} at different
radii. The parameters of the bar are, $\varepsilon_b=0.1$, $r_b=5$~kpc, and
$h=0.2$~Gyr. The lines are the same as in Fig.~\ref{fig::sfr_global}. Each
line corresponds to the SFR integrated over the area of a $1$~kpc width
ring centered on the value given in the legend of Fig.~\ref{fig::sfr_global}.}\label{fig::sfr_profiles}

\end{figure*}
%%%%%%%%%%%%%%%%%%%%%%%%%%%%%%%%%%%%%%%%%%%%%%%%%%%%%%%%%%%%%%

%%%%%%%%%%%%%%%%%%%%%%%%%%%%%%%%%%%%%%%%%%%%%%%%%%%%%%%%%%%%%%
\begin{figure*}
\includegraphics[width=0.49\hsize]{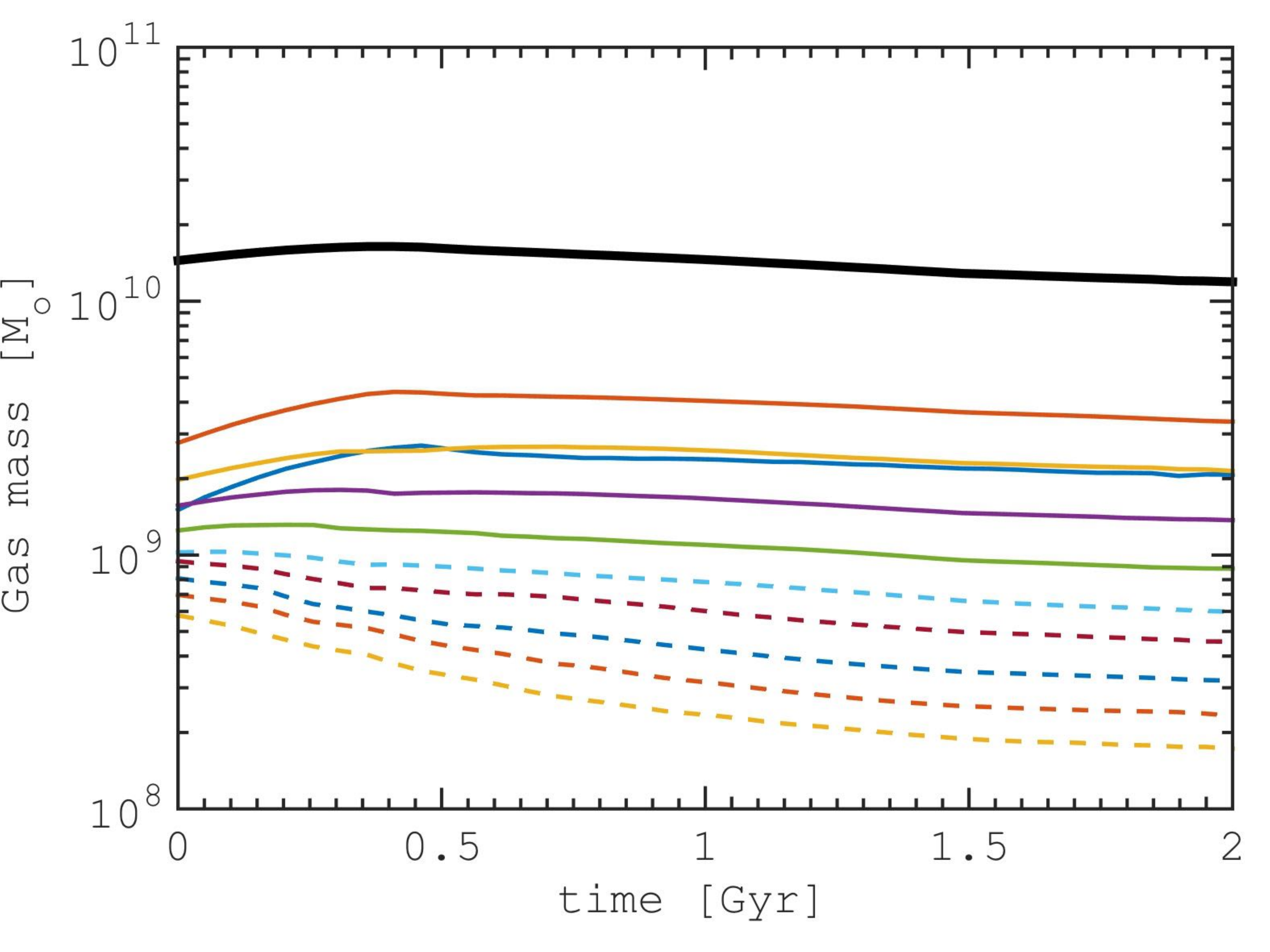}\hfill\includegraphics[width=0.49\hsize]{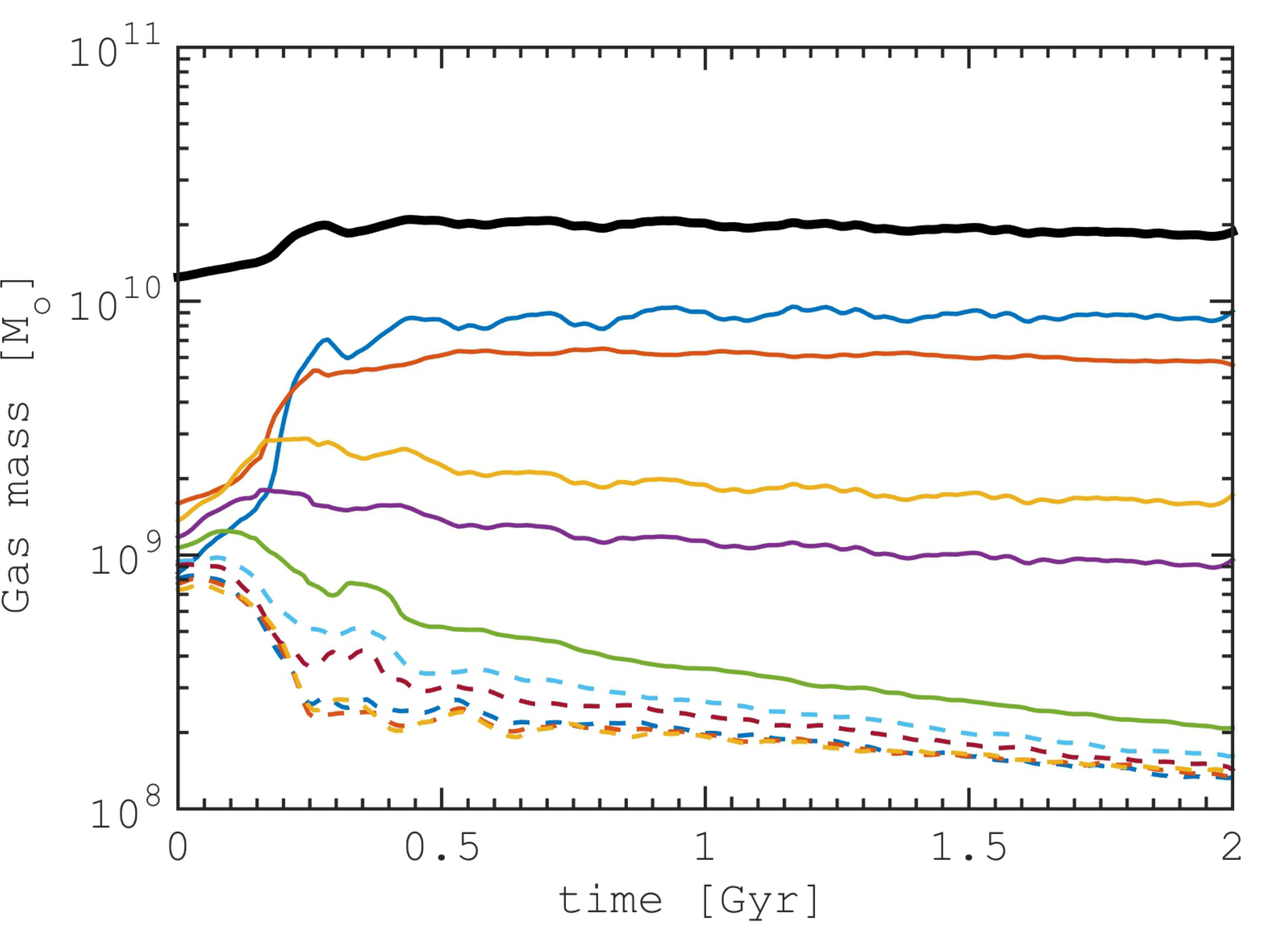}
\caption{The evolution of the mass of gas for the same models as show in Fig.~\ref{fig::sfr_profiles}. Here, the black
line corresponds to the total mass of gas with $15$~kpc radius of the disk.
}\label{fig::mass_profiles}
\end{figure*}

%%%%%%%%%%%%%%%%%%%%%%%%%%%%%%%%%%%%%%%%%%%%%%%%%%%%%%%%%%%%%%

The reason for the decrease in the star-formation ratio of the
barred versus unbarred models can be understood by comparing the
star-formation rate evolution at different disk radii in the two models
(see Fig.~\ref{fig::sfr_profiles}).  Since the star-formation rate depends
on the gas density, initially, $t=0$, it decreases from the center to the
outskirts of the disk. With time, the unbarred galaxy shows a monotonic
decrease of SFR at large radii ($r>5$~kpc), whereas the inner regions,
$r<5$~kpc, the SFR is nearly constant. That it is the amount of gas
available at a given radius the main driver of the temporal evolution
of the SFR in the unbarred galaxy (Fig.~\ref{fig::mass_profiles}).
Comparing Figs.~\ref{fig::sfr_profiles} and ~\ref{fig::mass_profiles}
for the unbarred model, one sees that the nearly constant SFR observed
in the inner regions of the unbarred disk is reflected into the nearly
constant gas mass for each region.

On the contrary, the SFR evolution in the barred galaxy is more
complex. After an initial peak, at about t=0.25~Gyr, just after the
growth of the bar has ended, the SFR generally decreases. Over
the innermost region, $r<3$~kpc, after an initial decrease,
the SFR is constant for $t > 0.7$~Gyr. At intermediate radii,
$3-7$~kpc, the star-formation rate decreases by a factor of $5-10$.
Star formation is mostly absent/suppressed within the bar radius
except for the strong gas concentration area along the bar major axis
(Fig.~\ref{fig::evolution_rigid_bar}). The overall SFR decreases by a
factor of about $10$. Comparing the evolution of the SFR and gas mass over
the same regions in the barred model, we see that the decrease in the SFR
observed in the inner regions is not due to a decrease in the gas mass.
Perhaps surprisingly, the decrease in the SFR occurs while there is
an increase in the gas content within about $3$~kpc of the galactic
center. This indicates that gas is losing angular momentum and flows
inwards accumulating in the center.  Once there, it is not converted
into  stars with the same efficiency observed for the unbarred model,
otherwise we would expect a corresponding increase of the SFR in the
center of the disk.

%%%%%%%%%%%%%%%%%%%%%%%%%%%%%%%%%%%%%%%%%%%%%%%%%%%%%%%%%%%%%%

\begin{figure*}
\includegraphics[width=0.49\hsize]{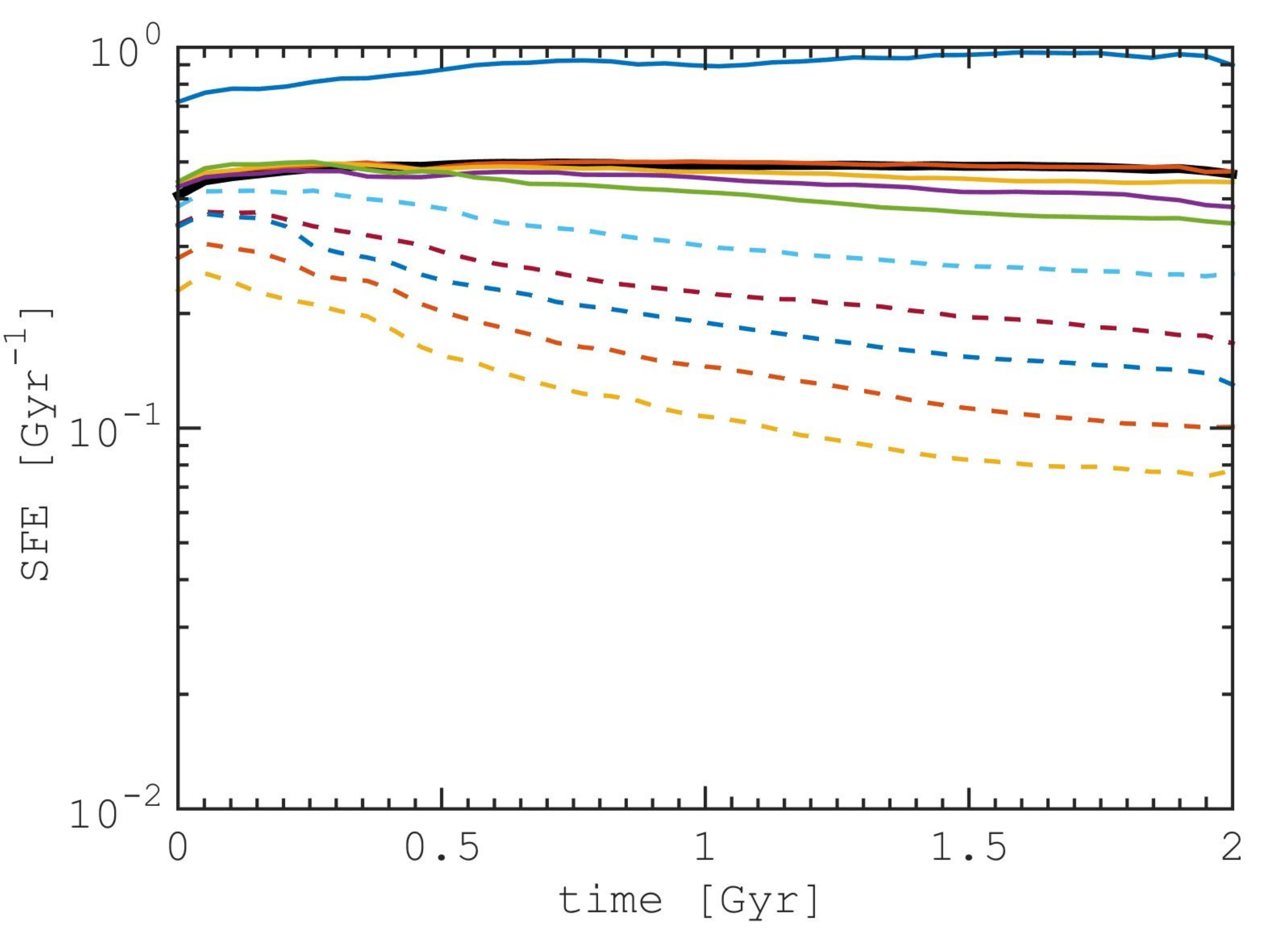}\hfill\includegraphics[width=0.49\hsize]{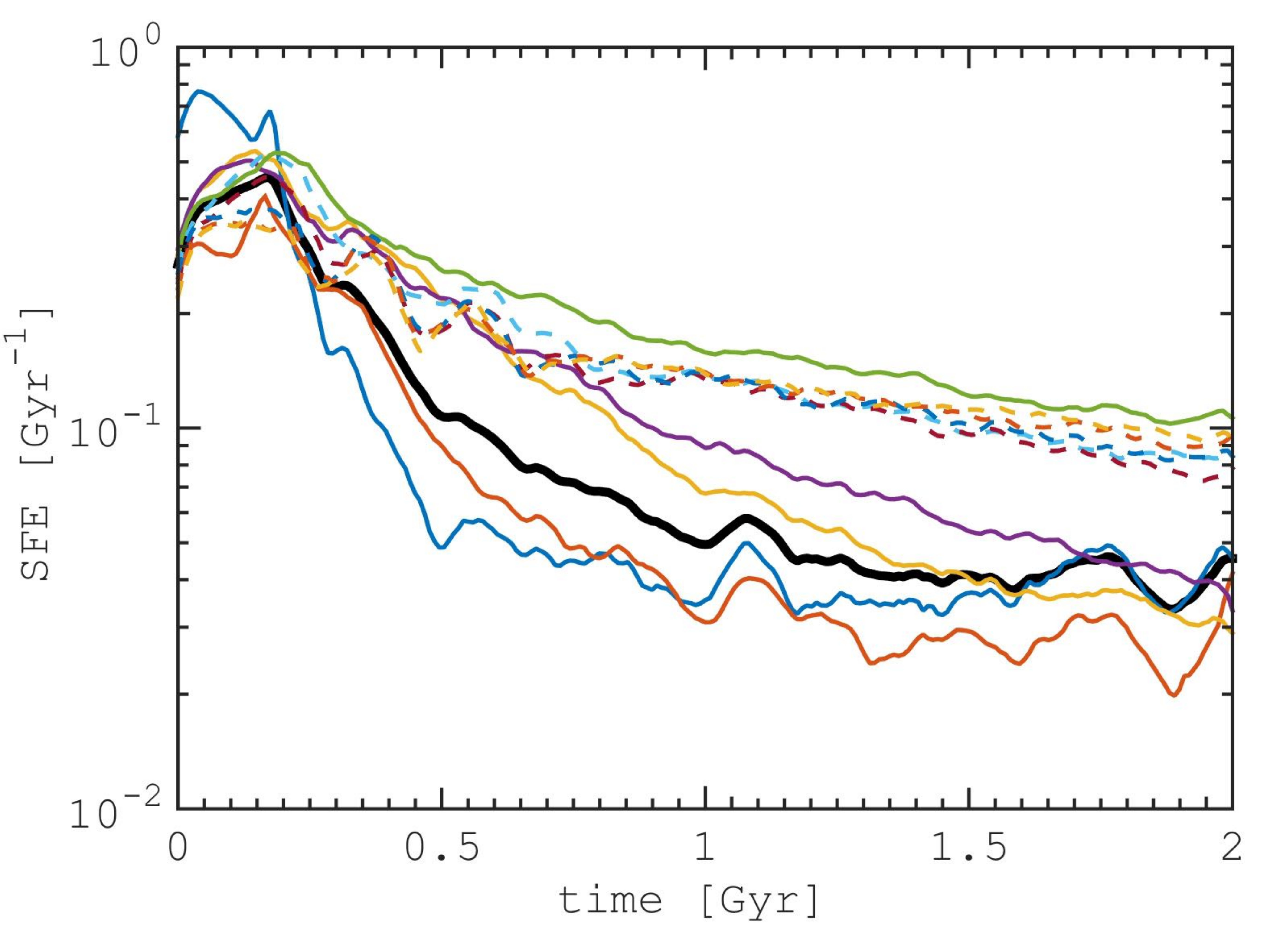}
\caption{The evolution of the star-formation efficiency,
$\Sigma_{\rm SFR}/\Sigma_{\rm gas}$, for the same models as in
Fig.~\ref{fig::sfr_profiles}.  In this figure, the black line corresponds
to the mean star-formation efficiency within 15 kpc of the center of
the disk.}
\label{fig::sfe_profiles}
\end{figure*}

%%%%%%%%%%%%%%%%%%%%%%%%%%%%%%%%%%%%%%%%%%%%%%%%%%%%%%%%%%%%%%

The evolution of the mass or surface density of gas at different radii
cannot explain quantitatively the decrease of the star-formation rate
in the barred galaxy. To emphasize this, we show the evolution of the
star-formation efficiency which is defined as the star-formation rate
surface density per unit gas mass surface density ${\rm SFE} = \Sigma_{\rm
sfr} / \Sigma_{\rm gas}$ with units of $yr^{-1}$. It is clear that
in the unbarred galaxy, the star-formation efficiency is roughly
constant in the inner regions, $r<5$, both in time and in space
(Fig.~\ref{fig::sfe_profiles}). It is only in the outer disk of the
unbarred model that the SFE decreases by a factor of a few. Whereas
for the barred galaxy, the star-formation efficiency decreases rapidly
at all radii by a factor of $2-10$ right after the formation of the
bar. Moreover, the SFE in the barred galaxy simulation is overall lower than it
is in the unbarred simulation.

It appears that the formation and presence of a bar strongly affects
the star-formation efficiency, especially within the bar scale
length ($r<5$~kpc in our models), by reducing it significantly.
So while gas that is shocked and dissipates angular momentum
can pile up at the center, it is not efficiently converted into
stars. In the centers of gas-rich barred galaxies, it appears that
an increase in the gas content is accompanied by a decrease in the
SFR.  It is widely accepted that there are several small and large
scale possible processes which  regulate star formation in galactic
disks~\citep{2007ARA&A..45..565M,2014prpl.conf....3D}. Gas random motions,
in particular, may provide an important mechanism counteracting gravity,
preventing the gas from becoming self-gravitating and from forming stars.

\subsection{Stellar bars and gas velocity dispersion}

%%%%%%%%%%%%%%%%%%%%%%%%%%%%%%%%%%%%%%%%%%%%%%%%%%%%%%
\begin{figure*}
\includegraphics[width=0.33\hsize]{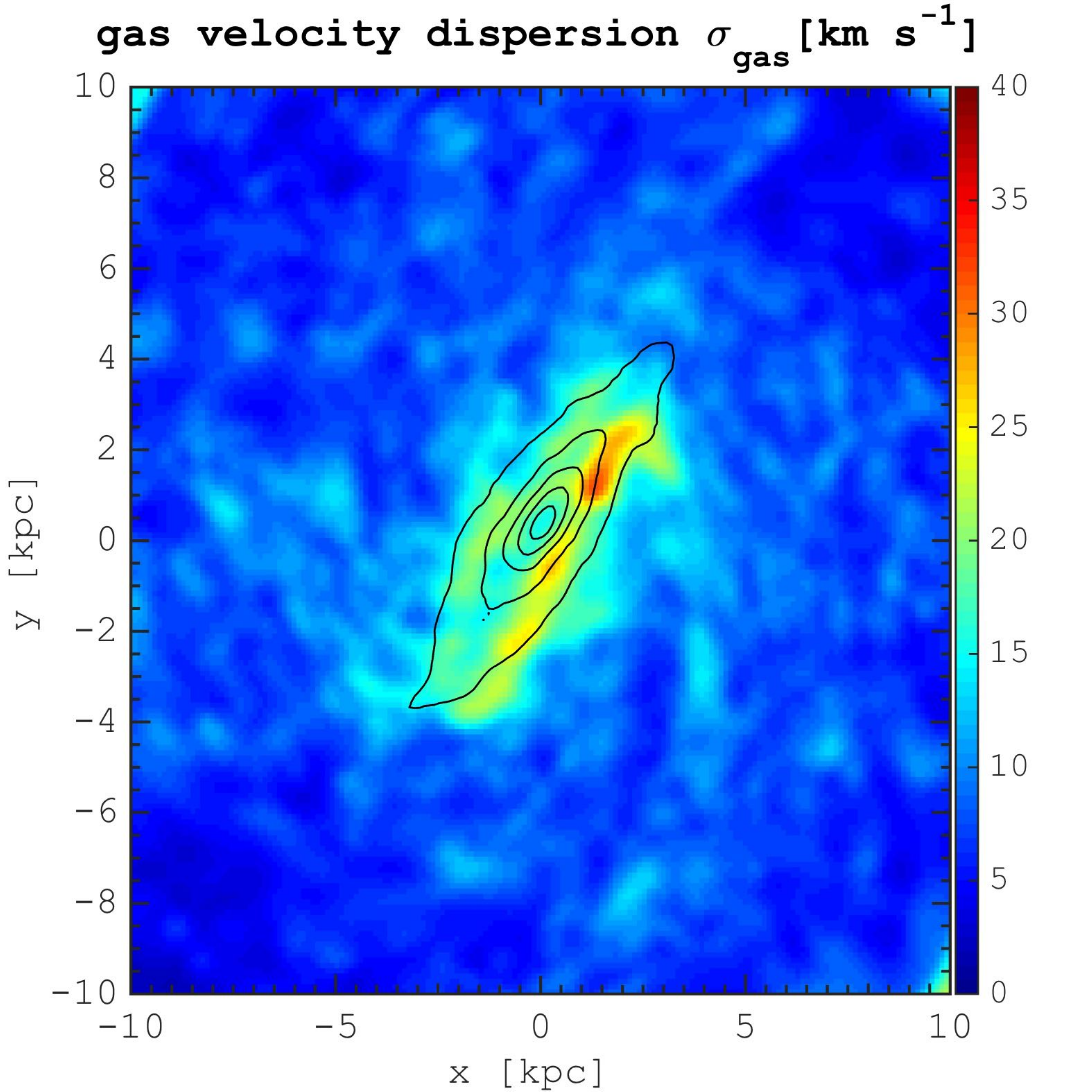}
\includegraphics[width=0.33\hsize]{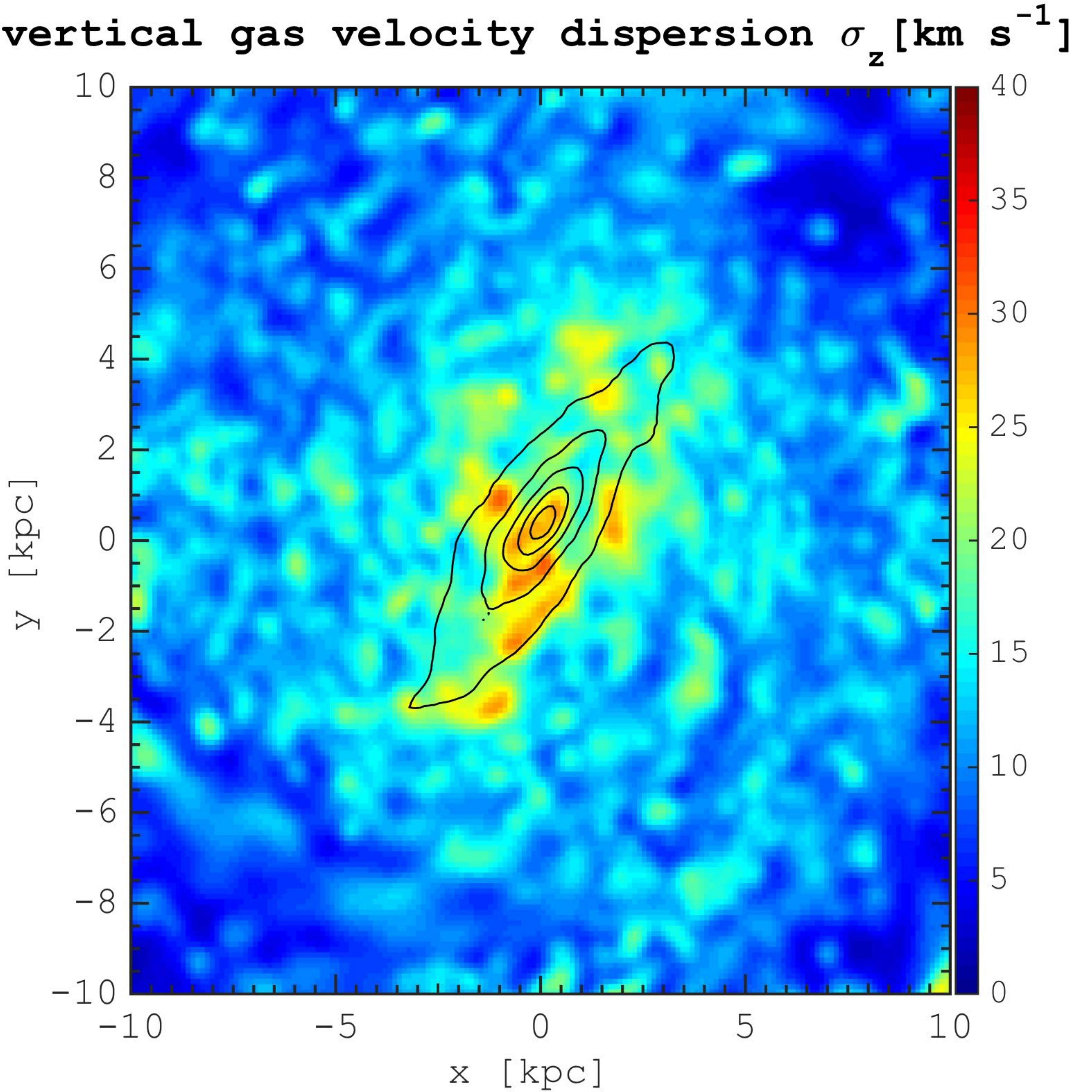}
\includegraphics[width=0.33\hsize]{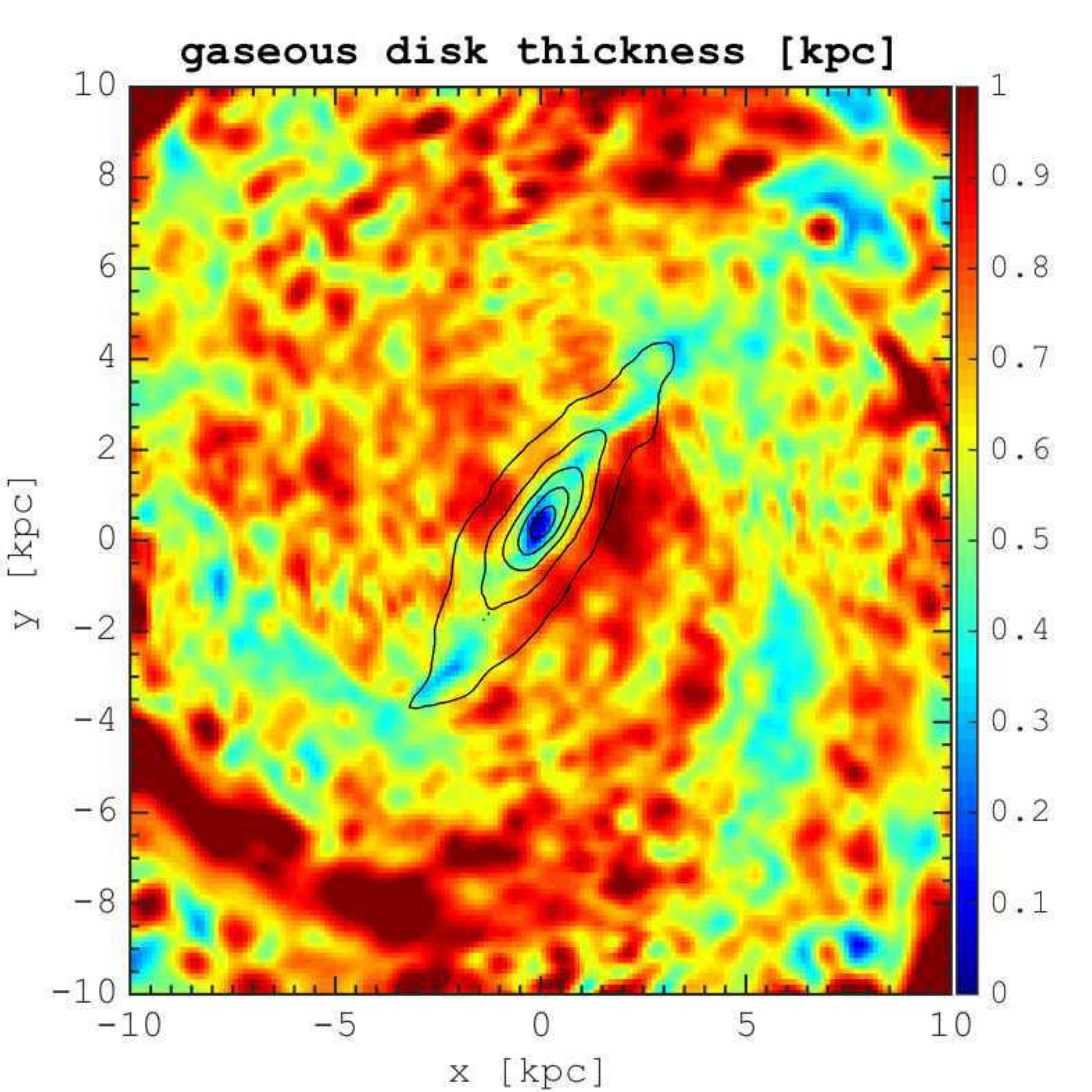}
\caption{Maps of the gas velocity dispersion $\sigma_{\rm gas}$ in
\kmps ~(left), vertical gas velocity dispersion $\sigma_{\rm z}$ in \kmps~(center) and gas disk thickness in kpc~(right) for barred galaxy model at a single time, $1.2$~Gyr.  Contours show the stellar surface density distribution in the vicinity of
bar.}\label{fig::dispersion_map}
\end{figure*}
%%%%%%%%%%%%%%%%%%%%%%%%%%%%%%%%%%%%%%%%%%%%%%%%%%%%%%

Mechanisms driving ISM turbulence in galaxies are still
not well-known or understood \citep[see][ and references
therein]{2016MNRAS.458.1671K}. Feedback from the recently formed stellar
populations is considered one of the most important mechanisms. It is
widely accepted that the large scale velocity dispersion correlates
with high SFR in disks \citep[see e.g.][]{2012MNRAS.421.3488H}, dwarf
galaxies \citep[see e.g.][]{2013MNRAS.432.1989S, 2015MNRAS.449.3568M}
and galaxies over a wide range of redshifts~\citep{2009ApJ...699.1660L,
2013A&A...555A..72L}. For instance, \cite{2006ApJ...638..797D} established
that a dispersion of $10$~\kmps can be sustained by a surface star
formation rate of about $(1-2)\times 10^{-3}$~\Msunkpc. There is
potentially enough energy and momentum in supernovae, radiative
feedback and stellar winds but it is not clear whether they are
deposited efficiently within gas in the disk. Circumstantial evidence
against stellar feedback as a direct source of turbulence is the lack
of correlation between the level of the velocity dispersion and the
proximity to star-forming clumps~\citep[see e.g.][]{2009ApJ...706.1364F,
2011ApJ...733..101G}.

On the other hand, local small scale self-gravity/shear due to galactic
rotation contribute to the formation of turbulent velocity fields
\citep[see e.g.][]{2009MNRAS.392..294A}. Galactic bar rotation can
supply an effectively inexhaustible amount of kinetic energy to power
turbulence in the ISM, where shocks can efficiently transform some of
the available bulk rotational energy into random gas motions. Since
most of the gas in the central part is pushed by the bar potential,
this gas can have a relatively large velocity dispersion which in
turn would affect the star-formation efficiency. The drivers of the
turbulent component of the velocity dispersion are gravity and shear. In
the presence of a bar, the leading shock waves swing in the center and
amplify, inducing gravitational torques in the gas and hence increasing gas streaming motions and
the local velocity dispersion \citep[see e.g.,][]{2002ApJ...581.1080K,
2007ApJ...660.1232K}. Such a process converts ordered circular motion
of the gas to random velocities, hence tapping rotational energy from
the disk. Numerical simulations clearly show that the stellar bar in
NGC~2915 enhances the ISM velocity dispersion~\citep{2002ApJ...577..197W}.

%%%%%%%%%%%%%%%%%%%%%%%%%%%%%%%%%%%%%%%%%%%%%%%%%%%%%%%%%%%%%%

\begin{figure*}
\includegraphics[width=0.49\hsize]{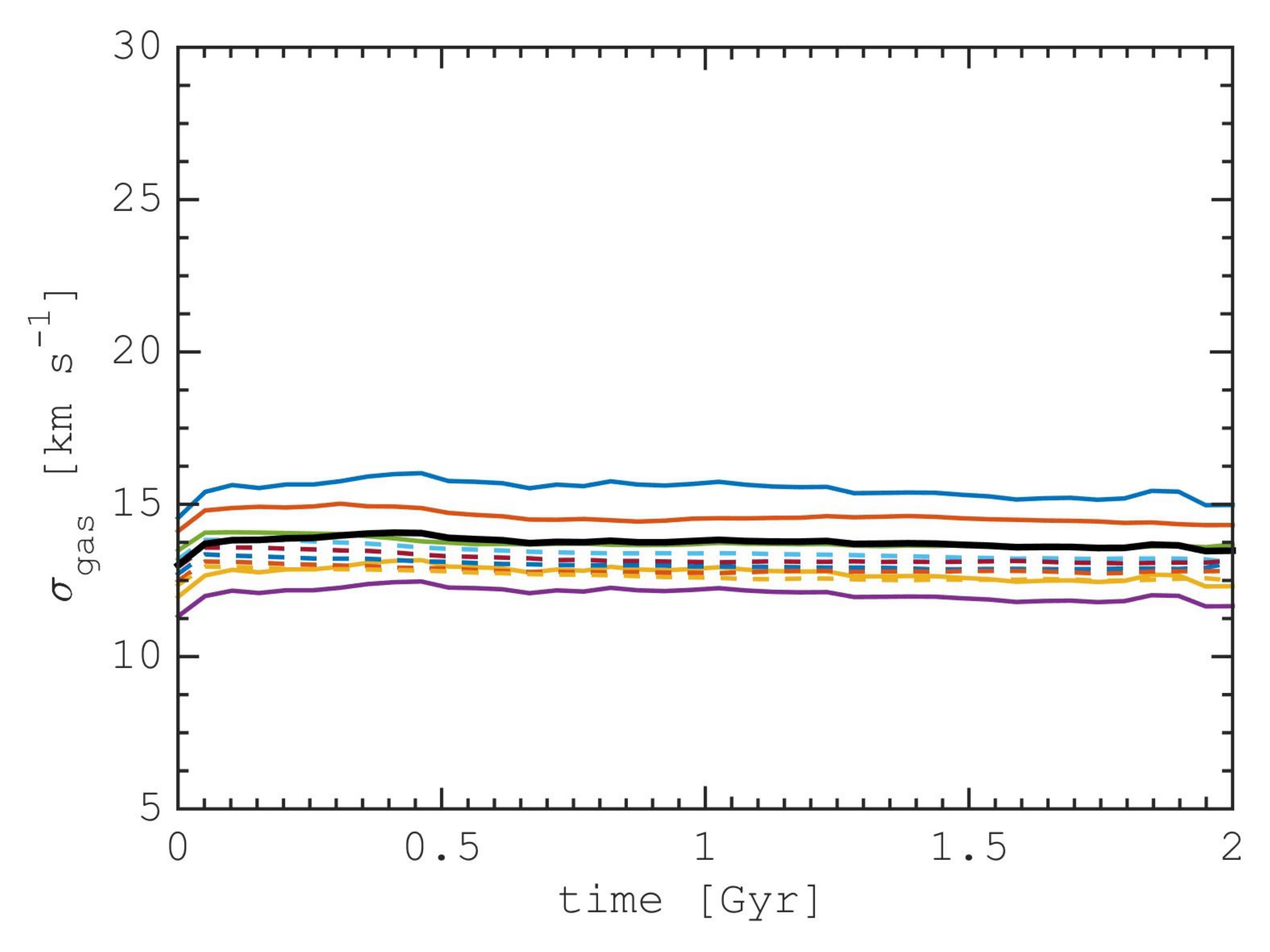}\hfill\includegraphics[width=0.49\hsize]{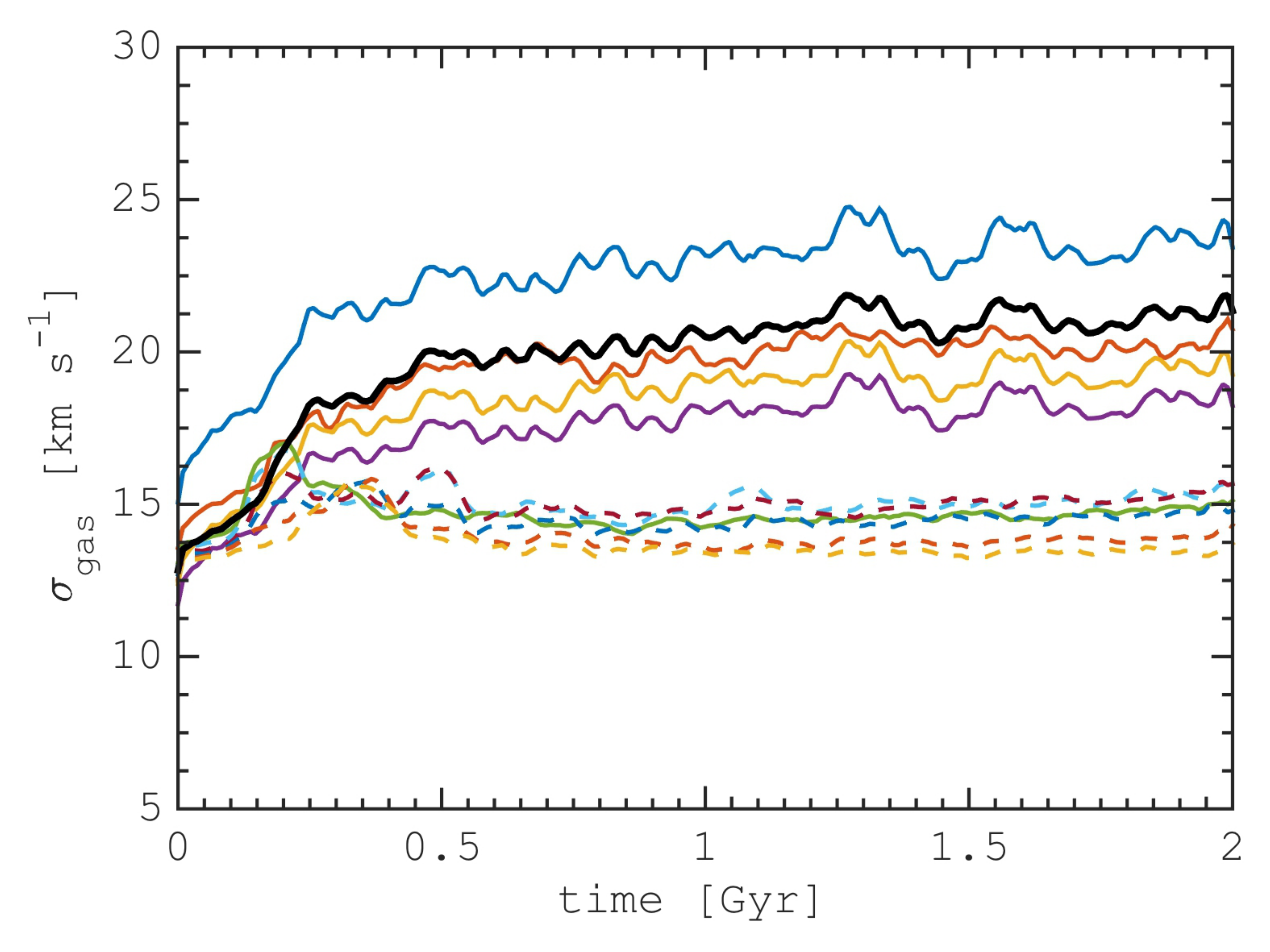}
\caption{The evolution of the gas velocity dispersion ~(see Eq.~\ref{eq::gas_disp}) for the same models as in Fig.~\ref{fig::sfr_profiles}.
The black line corresponds to the
mean gas velocity dispersion, $\sigma_{\rm gas}$, averaged over the inner
15~kpc of the disk.}\label{fig::dispersion_profiles}
\end{figure*}

%%%%%%%%%%%%%%%%%%%%%%%%%%%%%%%%%%%%%%%%%%%%%%%%%%%%%%%%%%%%%%
%\Oo (\sigma_{\rm gas})^2 =   \sum_{i=x,y,z} \langle(V_i(x,y) - \langle V_i(x,y)\rangle)^2 \rangle\,\label{eq::gas_disp}

Since there is a strong shear flow along the bar major axis, it is not
obvious how to distinguish this large scale gas flow from true random
motions. To check that the bar action provides an increase in
the gas {\it random} motions we calculate the gas velocity
dispersion for each two-dimensional position in the disk for the face-on galaxy configuration. We calculate the velocity dispersion for each component of the gas velocity~(radial, azimuthal and vertical) in columns across the disk plane. Then we calculate the mean gas velocity dispersion for each position $x,y$ in face on map:
\begin{equation}
\Oo \sigma_{\rm gas} = \sqrt{ \frac{1}{3} \left(\sigma^2_{\rm R} + \sigma^2_{\rm \phi} + \sigma^2_{\rm z} \right)} \,\label{eq::gas_disp}
\end{equation}
where $\sigma_{\rm R}$, $\sigma_{\rm \phi}$, $\sigma_{\rm z}$ are the velocity dispersions in the various directions. In Fig. \ref{fig::dispersion_map} we
plot the total gas velocity dispersion in the barred galaxy simulation. We
find that the bar region is characterized by a high gas velocity
dispersion, which can locally reach up to $25-35$~\kmps.
To check the role of the gas velocity dispersion definition, we also compare our definition Eq.~\ref{eq::gas_disp} with a simple line-of-sight gas velocity dispersion for the same face-on galaxy configuration which is simply the vertical gas velocity dispersion $\sigma_{\rm z}$. In Fig.~\ref{fig::dispersion_map}~(center) we can clearly see that there is nice spatial correlation between $\sigma_{\rm z}$ and $\sigma_{\rm gas}$. We also see that higher vertical gas velocity dispersion coincides well with the bar position. However, the most important consequence of this picture is the relative increase of the gas random motions in the bar region in comparison to unbarred configuration.

For the model without bar, the gas velocity dispersion $\sigma_{\rm
gas}$ is supported only by the stellar feedback and $\sigma_{\rm gas}$
does not exceed $15$~\kmps~(Fig~\ref{fig::dispersion_profiles})
which is in a agreement with a number of studies~\citep[see
e.g.][]{2013AJ....146..150C}. Usually gas velocity dispersion is
assumed to be equal to $11$~\kmps which is a typical value for
the HI gas in a local sample of galaxies \citep[the THINGS sample;
][]{2009AJ....137.4424T}. Note, however, that we have a relatively high
star-formation rate in the model and our definition of $\sigma_{\rm gas}$
cannot be directly compared with these observational numbers because we
do not consider both the thermal and kinetic components of the observed
velocity dispersion.

For the model with bar, the gas velocity dispersion increases
rapidly after $0.2-0.5$~Gyr, right after the bar amplitude becomes
substantial. Mean values of $\sigma_{\rm gas}$ remain higher in the
central $5$~kpc, up to $25$~\kmps. In particular, the gas velocity
dispersion $\sigma_{\rm gas}$ has a maximum along the bar major axis
peaking up to $\approx35$~\kmps. In the outer parts, the gas velocity
dispersion is constant and agrees with the corresponding values in the
unbarred galaxy simulations.

\subsection{Gaseous disk thickness and its stability}

%%%%%%%%%%%%%%%%%%%%%%%%%%%%%%%%%%%%%%%%%%%%%%%%%%%%%%%%%%%%%%

\begin{figure*}
\includegraphics[width=0.33\hsize]{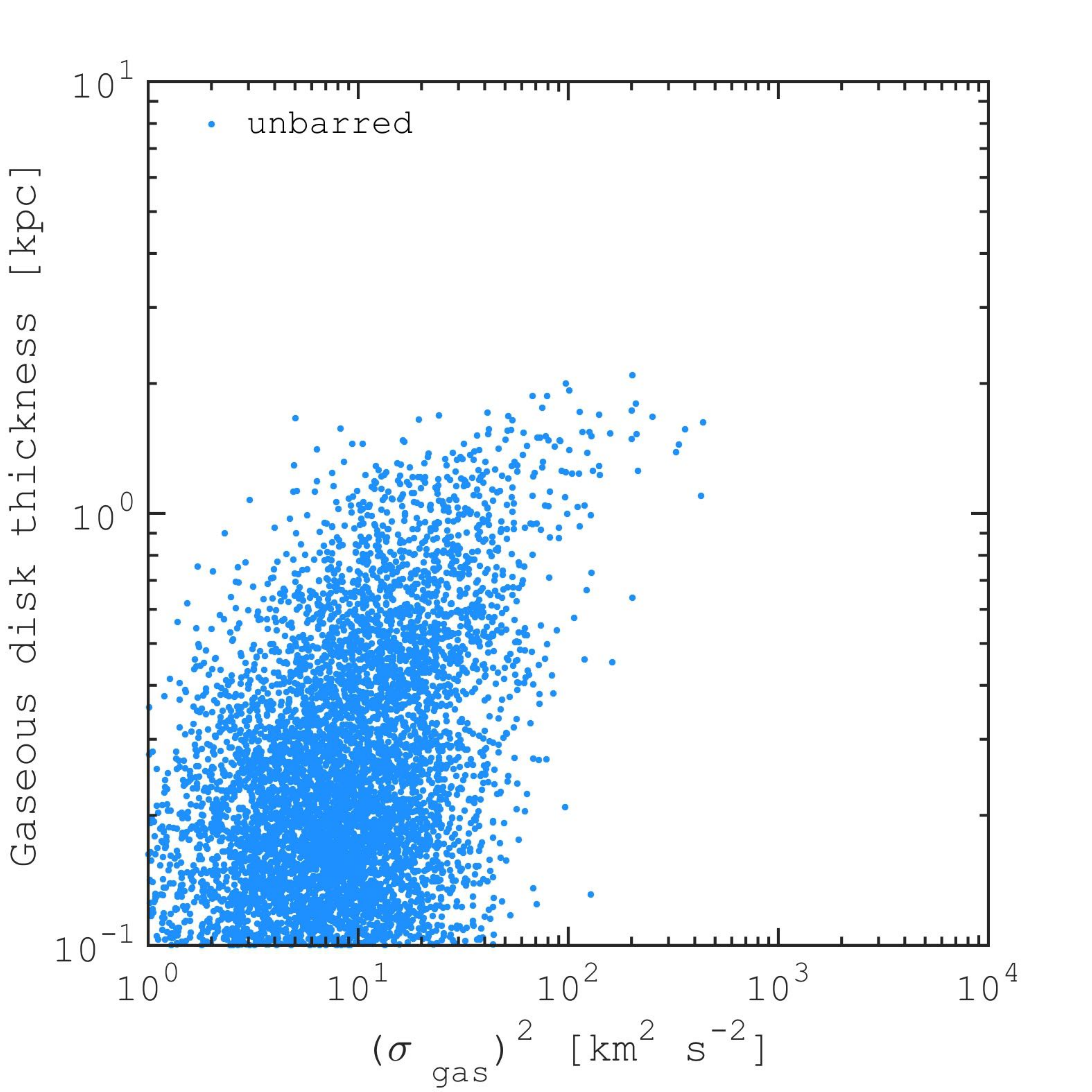}\includegraphics[width=0.33\hsize]{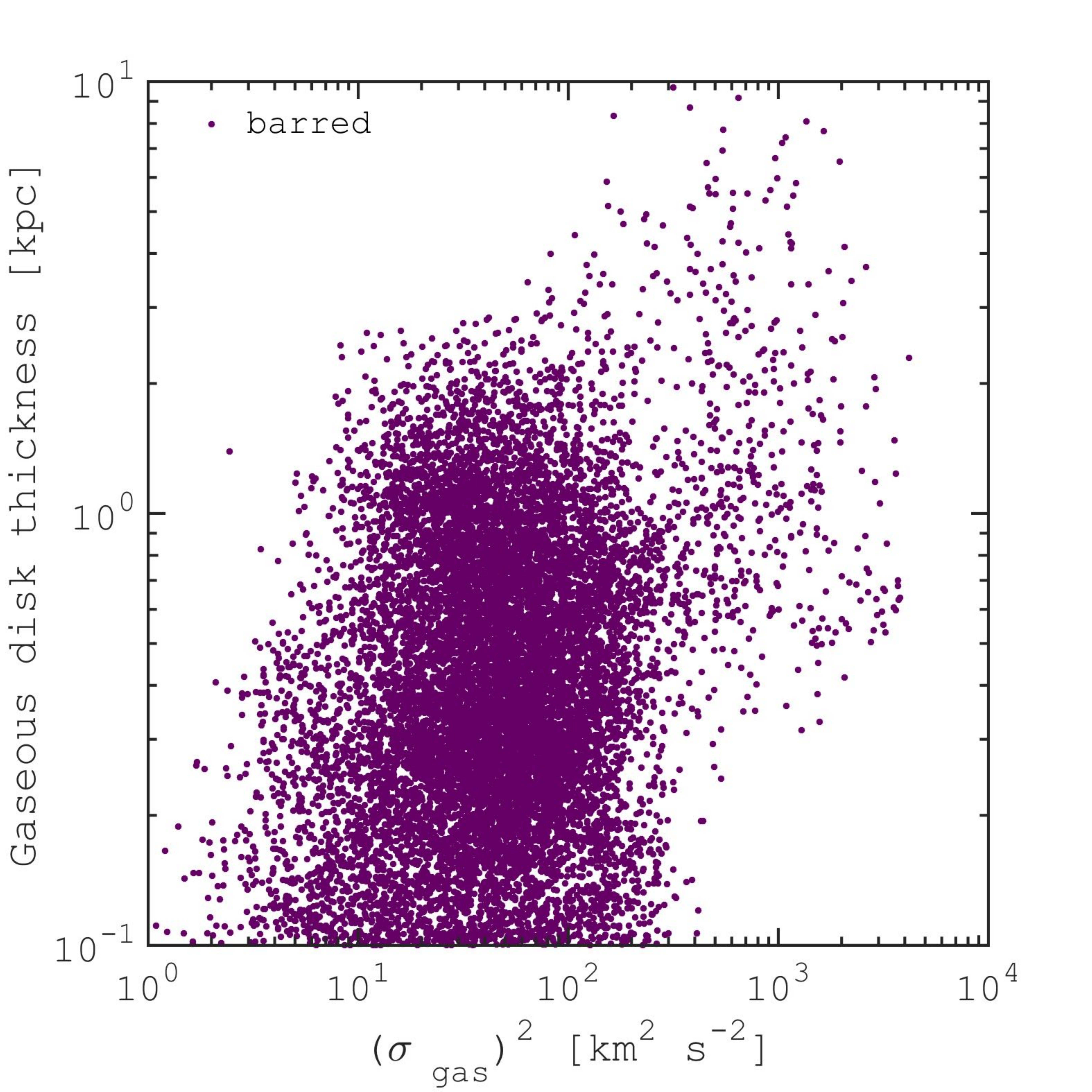}\includegraphics[width=0.33\hsize]{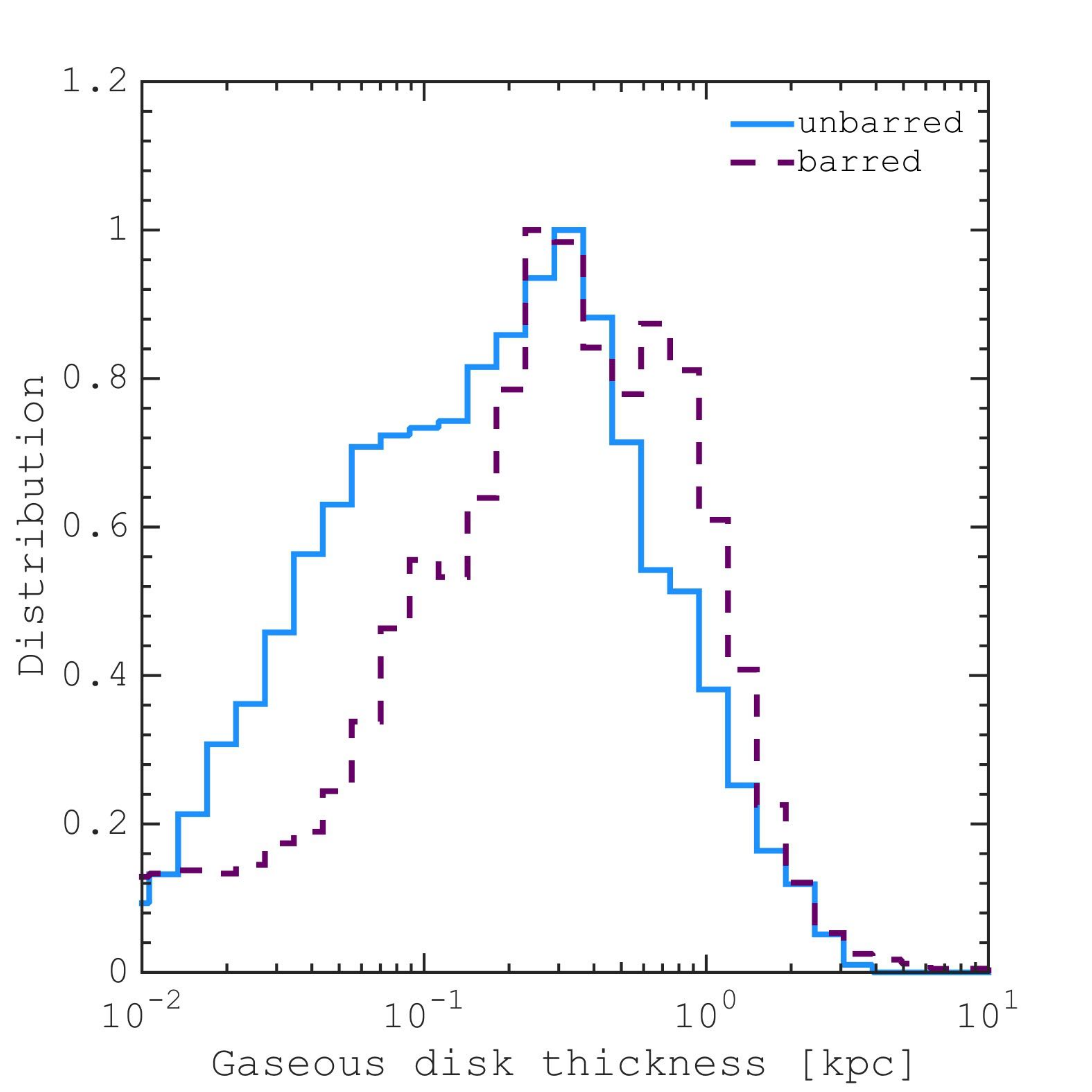}
\caption{Relation between the gaseous disk thickness and the gas velocity dispersion for the unbarred galaxy simulation~(RA) and the barred galaxy model~(RB). This relation is shown for values within the inner $5$~kpc of the disk and at $1.2$~Gyr in each simulation \textit{(left and center panels)}. \textit{Right panel:} The distribution
functions of the thickness of the gaseous disk for the two models shown in the left and center panels.}\label{fig::gas_thickness}
\end{figure*}

%%%%%%%%%%%%%%%%%%%%%%%%%%%%%%%%%%%%%%%%%%%%%%%%%%%%%%%%%%%%%%

Since star formation takes place in gravitationally unstable
galactic disks, it can be quenched when a disk becomes stable
against fragmentation. It is widely accepted that galactic
disk stability is affected by several factors including the
vertical structure of the disk and gas turbulence \citep[see
e.g.,][]{2011MNRAS.416.1191R, 2011ApJ...737...10E, 2012MNRAS.425.1511H}. 
The effect of disk thickness is to increase the effective
stability parameter \citep[e.g., Eqs. 15, 19 in][]{2013MNRAS.433.1389R} --
an increase in the thickness of either the gaseous or stellar disks
stabilizes such a two component galactic
disk. Disk stability analyses usually assumes an
axisymmetric galaxy disk which can not be globally explored in the
case of non-axisymmetric configurations, such as that in bar-dominated
galaxies. However, we try to address the effect induced by the growth
of the gas velocity dispersion on the disk thickening in the following.

In equilibrium, for a given gravitational potential, axisymmetric gaseous
disk thickness is determined by the gas velocity dispersion, $h_{\rm
gas} \propto \sigma^2_{\rm gas}$. In the previous section we showed that
the action of the stellar bar is to efficiently increase the gas random
motions. Thus we naturally expect a thickening of the gaseous disk due to
its higher velocity dispersion in the bar region, which is seen in Fig.~\ref{fig::dispersion_map} for $r<5$~kpc. To demonstrate this point quantitatively we show the relation between disk thickness, defined as
the mean absolute value of the mass weighted vertical position of gas,
versus square of gas velocity dispersion~(see Eq.~\ref{eq::gas_disp})
for unbarred~(RA) and barred galaxy simulation~(RB) at $1.2$~Gyr
and within inner 5~kpc (Fig.~\ref{fig::gas_thickness}). For the unbarred
galaxy, the gas velocity dispersion is mostly induced by the stellar
feedback which is rather high at the current level of the star formation,
$\approx 8$~\Msunyr. For the barred galaxy, the gaseous disk thickness
tends to be larger compared to unbarred model, with significantly high
values of thickness~($>1$~kpc) for the highest velocity dispersions.
These high velocity dispersion regions are
within the bar region~(Fig.~\ref{fig::dispersion_map}).

To derive a quantitative difference between the disk thickness in the
barred and unbarred galaxies, we show the disk thickness distribution
functions (Fig.~\ref{fig::gas_thickness}). In both models the distribution
functions are very wide. The mean value of the thickness for the unbarred
galaxy is close to $0.2$~kpc while for the barred galaxy there are two
peaks in the distribution function. The lowest peak is very close to the
one for the observed for the unbarred galaxy. The highest peak corresponds
to r$\approx 0.8$~kpc which is the manifestation of bar-induced disk
thickening. Thus we can  argue that stellar bar action and the following
gas velocity dispersion growth lead to the disk thickening which is
important stabilizing the two-component disk.

A detailed stability analysis is complicated due to strongly
non-axisymmetric configuration of our models with bars. In lieu of a
detailed stability analysis, we simply suggest that the growth of the
gas velocity dispersion and gaseous disk thickness tends to stabilize the
galactic disk against gravitational fragmentation which in turn suppresses
the star formation. From this point of view, our results are in agreement
with the mechanism of morphological quenching \citep{2009ApJ...707..250M}.
In \cite{2009ApJ...707..250M}, the disk is stabilized due to the growth of
the dynamically hot spheroid which is built up by successive mergers. In
another words, the quenching mechanism is the result of the increase of
the stellar velocity dispersion induced by interactions.  In this study,
we propose a mechanism for the stabilization of the gaseous disk that
does not need environment to play any role. In one sense, both studies
reach the same conclusion, increasing the velocity dispersion of one
components, stars or gas, stabilizes the disk against fragmentation and
thus suppresses star formation.

\subsection{The $\Sigma_{\rm SFR} - \Sigma_{\rm gas}$ relation}

\begin{figure*}
\includegraphics[width=1\hsize]{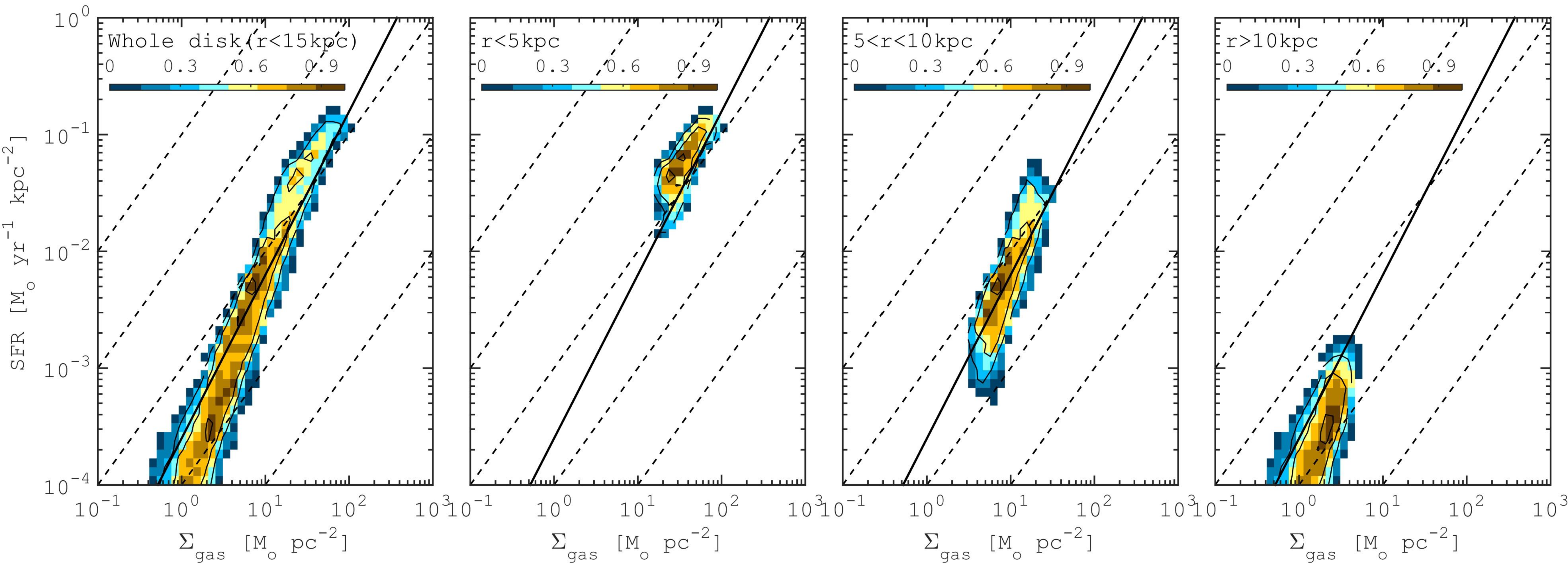}\\\includegraphics[width=1\hsize]{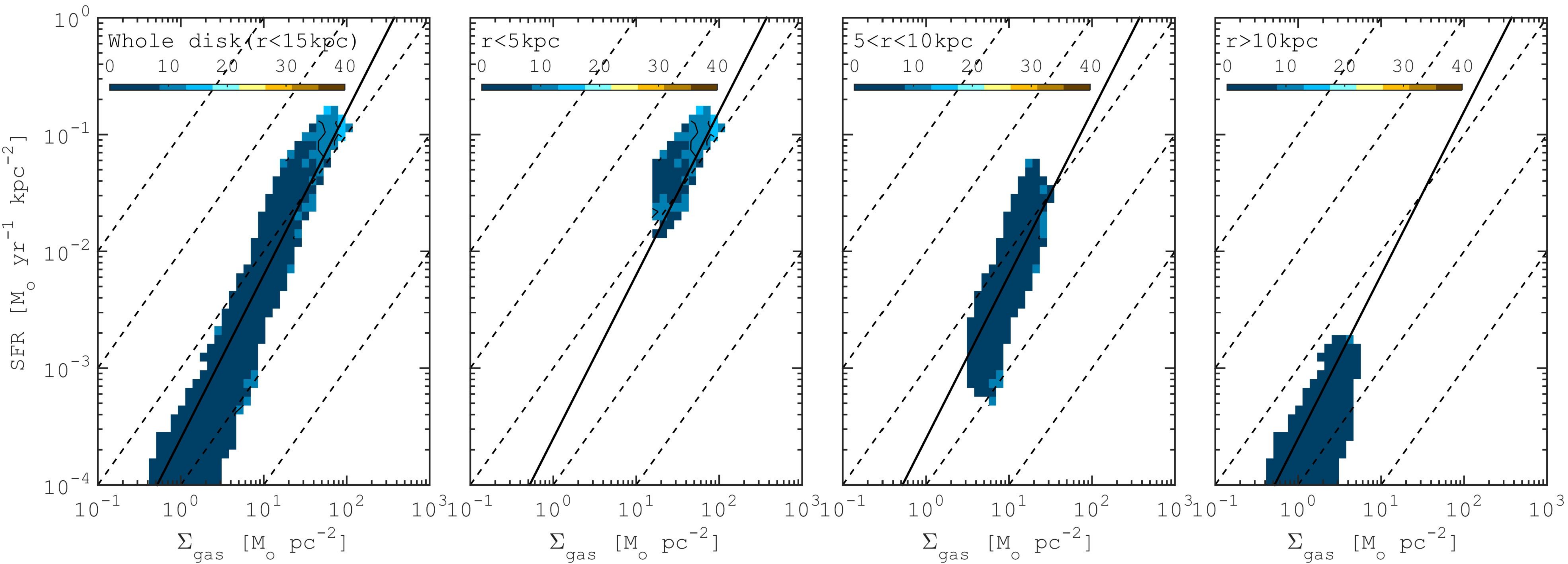}
\caption{Relationship between the star-formation rate surface density
and the gas surface density, $\Sigma_{\rm SFR} - \Sigma_{\rm gas}$,
in the model without bar~(RA) at a single time, $1.2$~Gyr for a variety of
spatial scales over the disk. In the {\it top row} color represents
normalised density (from $0$ to $1$) with a given SFR and gas density;
{\it in bottom row} color represents a mean gas velocity dispersion and
contours show the values from $10$~\kmps to $40$~\kmps with a bin size
of $5$~\kmps.  \textit{Panels from left to right:} The K-S relationship
using data from the entire disk; from the inner, $r<5$~kpc; from
the radii between $5<r<10$~kpc; and from the outer parts of the disk,
$r>10$~kpc. The solid black line represents the K-S relation with a slope
of $1.4$~\citep{1998ApJ...498..541K}. The diagonal dashed lines cutting
across each panel represents the star-formation efficiencies of $100\%$,
$10\%$, $1\%$, $0.1\%$ and $0.01\%$ per $10^9$~yr.}
\label{fig::ks_nobar}

\end{figure*}

Another way to understand the significance of the interplay between gas
content, star-formation rate and gas velocity dispersion is to investigate
the star-formation rate surface density versus the gas surface density
(Fig.~\ref{fig::ks_nobar}). We choose to illustrate this at a particular
moment in the simulation, $1.2$~Gyr, because at this time the quenching
episode occurred quite recently.  The star-formation rate in a given
model element is represented by the mass of stars formed during the
previous $10$~Myr and the gas surface density is taken from the single
snapshot at $t=1.2$~Gyr. 

For the unbarred galaxy, the $\Sigma_{\rm SFR} - \Sigma_{\rm
gas}$ relation seems to be in agreement with the
classical Kennicutt-Schmidt law \citep[K-S,][]{1959ApJ...129..243S,
1998ApJ...498..541K}. Indeed, depletion timescale increases with radius,
from a few Gyrs for the center to $10$~Gyr in the outskirts. Central star
formation rate is close to the intensity of starburst galaxies~\citep[see
e.g.,][]{1998ApJ...498..541K}, because the maximum of the star
formation occurs in the central region where the SFR is in the range
$6-8$~\Msunyr~(see Fig.~\ref{fig::mass_profiles}). Meanwhile, much
less efficient outer star formation is similar to those found in nearby
galaxies \citep{2010AJ....140.1194B}.

%%%%%%%%%%%%%%%%%%%%%%%%%%%%%%%%%%%%%%%%%%%%%%%%%%%%%%%%%%%%%%

%
%
%%%%%%%%%%%%%%%%%%%%%%%%%%%%%%%%%%%%%%%%%%%%%%%%%%%%%%%%%%%%%%

%%%%%%%%%%%%%%%%%%%%%%%%%%%%%%%%%%%%%%%%%%%%%%%%%%%%%%%%%%%%%%

\begin{figure*}
\includegraphics[width=1\hsize]{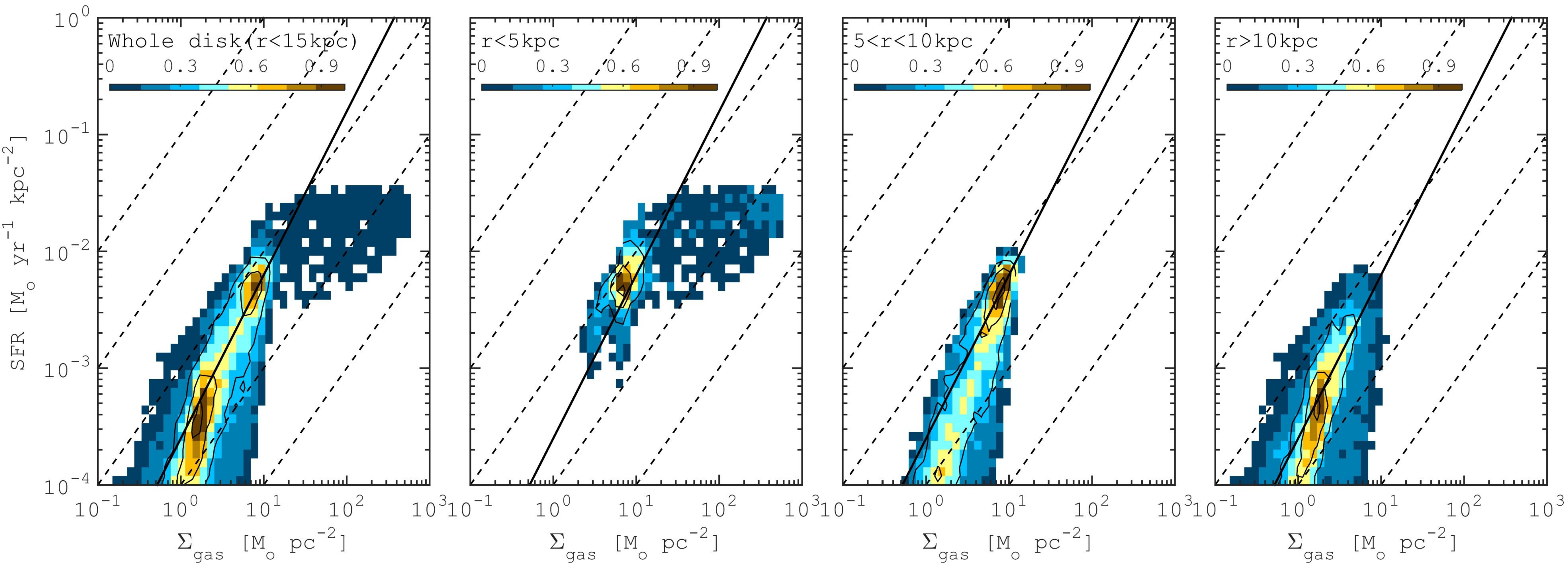}\\
\includegraphics[width=1\hsize]{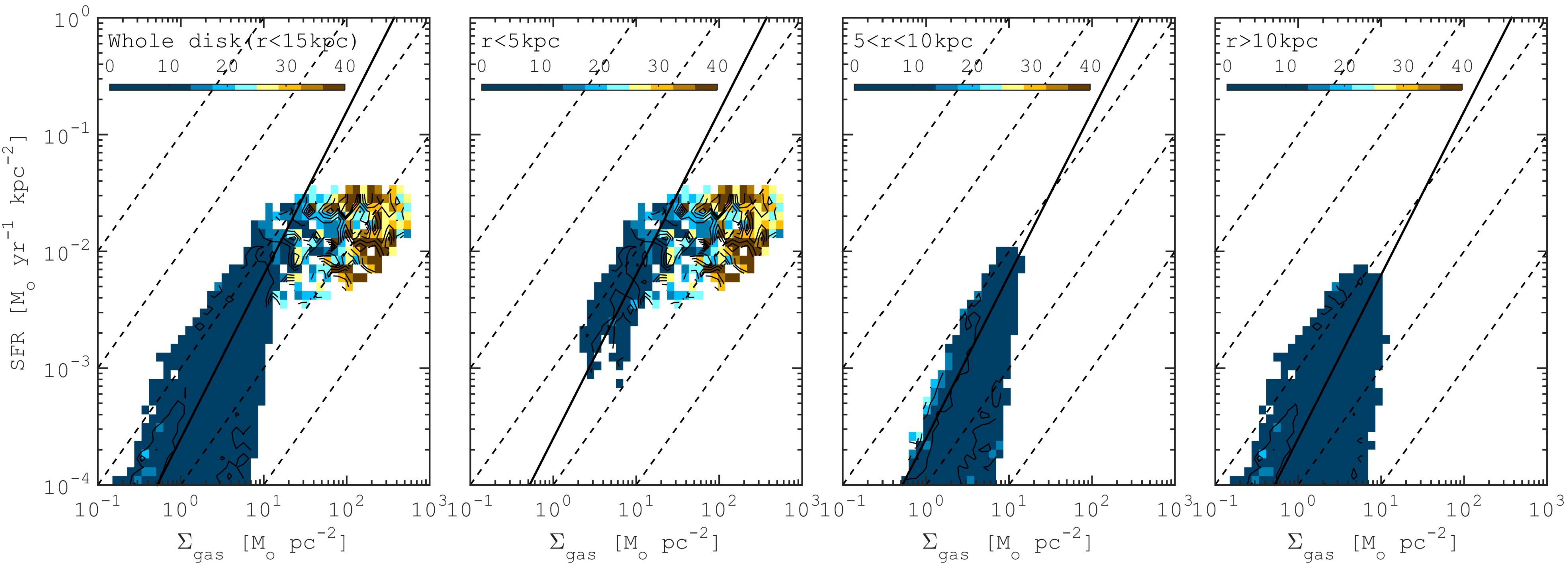}
\caption{Same panels as in Fig.~\ref{fig::ks_nobar}, but now
for the model containing a bar~(RA). The contours and lines are also the same
as in Fig.~\ref{fig::ks_nobar}.}\label{fig::ks_bar}
\end{figure*}

%%%%%%%%%%%%%%%%%%%%%%%%%%%%%%%%%%%%%%%%%%%%%%%%%%%%%%%%%%%%%%

We have already shown that there is a variation of the star-formation
efficiency with radius and time, most significantly in the barred
galaxy model. We showed that averaged SFE in the bar region decreases
by a factor of $2-10$ in comparison to the unbarred galaxy simulation
(Fig.~\ref{fig::sfe_profiles}). When the disk is quenching, the
Kennicutt-Schmidt like relation is also different from the behavior
observed in the unbarred galaxy. Figure~\ref{fig::ks_nobar} shows the
relation $\Sigma_{\rm SFR} - \Sigma_{\rm gas}$ at $1.2$~Gyr calculated
for the barred galaxy. The barred galaxy simulation contains the same
amount of gas as the unbarred one, whereas non-axisymmetric bar action
results in much higher densities. There is also a clear variation in
the SFE with radius that is overall similar to the unbarred galaxy
(Fig. \ref{fig::sfe_profiles}), with a decrease by a factor of $10$
from the inner to the outer disk. The bar has little impact on the outer
disk as the SFE is already low there.

Figure~\ref{fig::ks_bar} shows clear evidence for a break in the SFE
in the central region where the gas velocity dispersion is very high
(Fig.~\ref{fig::dispersion_map}). Most pixels agree well with K-S-type relation meanwhile the filling factor of suppressed star formation regions is low.  At the same time, the K-S relation for
the middle and outer parts of the disk is roughly the same for barred
and unbarred galaxy models. In the densest regions which lie along the
bar major axis and in the central over density and where the gas surface
density can reach up to $10^3$~\Msunpc, we find a very wide range of
star-formation efficiencies. The SFE of regions where there is very
dense gas is significantly reduced, which explains the overall lower
star-formation rate.  However, the high amount of gas still generates a
higher star-formation rate than in the outer disk, where star formation
always contributes only a small fraction of the total SFR.

%%%%%%%%%%%%%%%%%%%%%%%%%%%%%%%%%%%%%%%%%%%%%%%%%%%%%%%%%%%%%%

\begin{figure*}
\includegraphics[width=1\hsize]{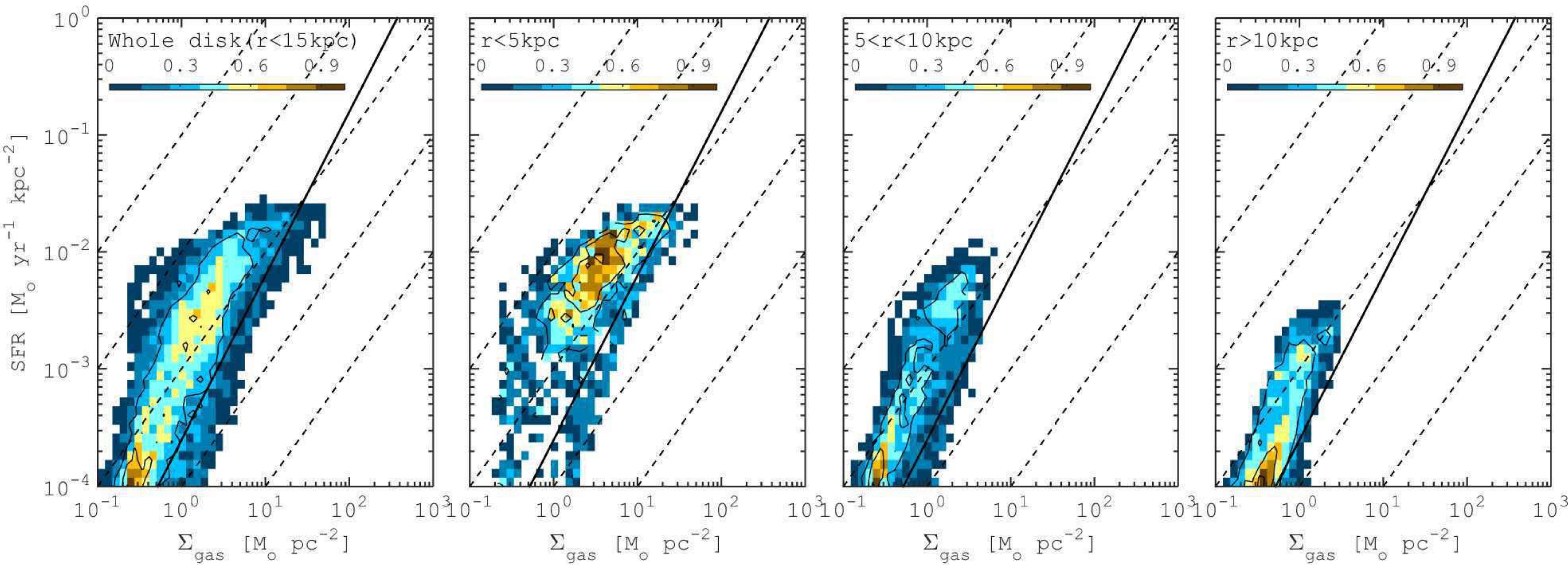}
\caption{Same panels as in Fig.~\ref{fig::ks_bar}~(top), but now
for the model containing a bar and a low gas fraction~(RBM02). The contours and lines are also the same
as in Fig.~\ref{fig::ks_bar}.}\label{fig::ks_barlowgasfraction}
\end{figure*}

%%%%%%%%%%%%%%%%%%%%%%%%%%%%%%%%%%%%%%%%%%%%%%%%%%%%%%%%%%%%%%

As previously described, Fig.~\ref{fig::ks_bar} shows the flattening of the K-S relation at high densities for the gas-rich model.  In Fig.~\ref{fig::ks_barlowgasfraction} we plot the K-S relation for a simulation with a gas fraction of  $10\%$. Also for this model, the slope of the K-S relation is $\approx 1.4$ in the disk, while in a limited region at high densities the relation appears to have a flatter slope, of $\approx 1$. So, while a hint of a flattening in the K-S relation is visible also for the barred, gas-poor galaxy, it is clear that the effect is much stronger in gas-rich systems. Currently, only few studies have compared the SFE in the bar region and outside it. They generally reach the conclusion that SFEs in bars are lower than in the surrounding disk. \cite{2010ApJ...721..383M}, for example, have studied NGC 4303, a barred galaxy, and shown that the K-S relation appears flatter in the bar region than in the surrounding disk. They show indeed that there is active star formation in the spiral arms, while  in the bar region the SFE is lower by a factor of $\approx 2$. For the barred galaxy NGC~1530, \cite{1998A&A...337..671R} claimed that the star formation is inhibited at the places where the shocks and the shear are strong enough. By making an analysis of radial distributions of the different types of supernovae  \cite{2016MNRAS.456.2848H} reported about the substantial suppression of massive star formation in the radial range swept by strong bars. 

In Maffei~2 \cite{2012PASJ...64...51S} found that molecular gas in the bar ridge regions is gravitationally unbound, which suggests that it can hardly become dense enough to form stars.  Such a gravitationally unbound condition may decrease the star-formation efficiency in the bar region. Note however, that none of these studies concern galaxies at high redshift or particularly gas-rich. Here we suggest that the mechanism of the bar-induced  quenching is efficient in gas-rich galaxies and the flattening of the K-S relation in the central regions may also be considered as a prediction to be checked with observations of high-redshift, or local gas-rich galaxies.

%%%%%%%%%%%%%%%%%%%%%%%%%%%%%%%%%%%%%%%%%%%%%%%%%%%%%%
\begin{figure*}
\includegraphics[width=0.33\hsize]{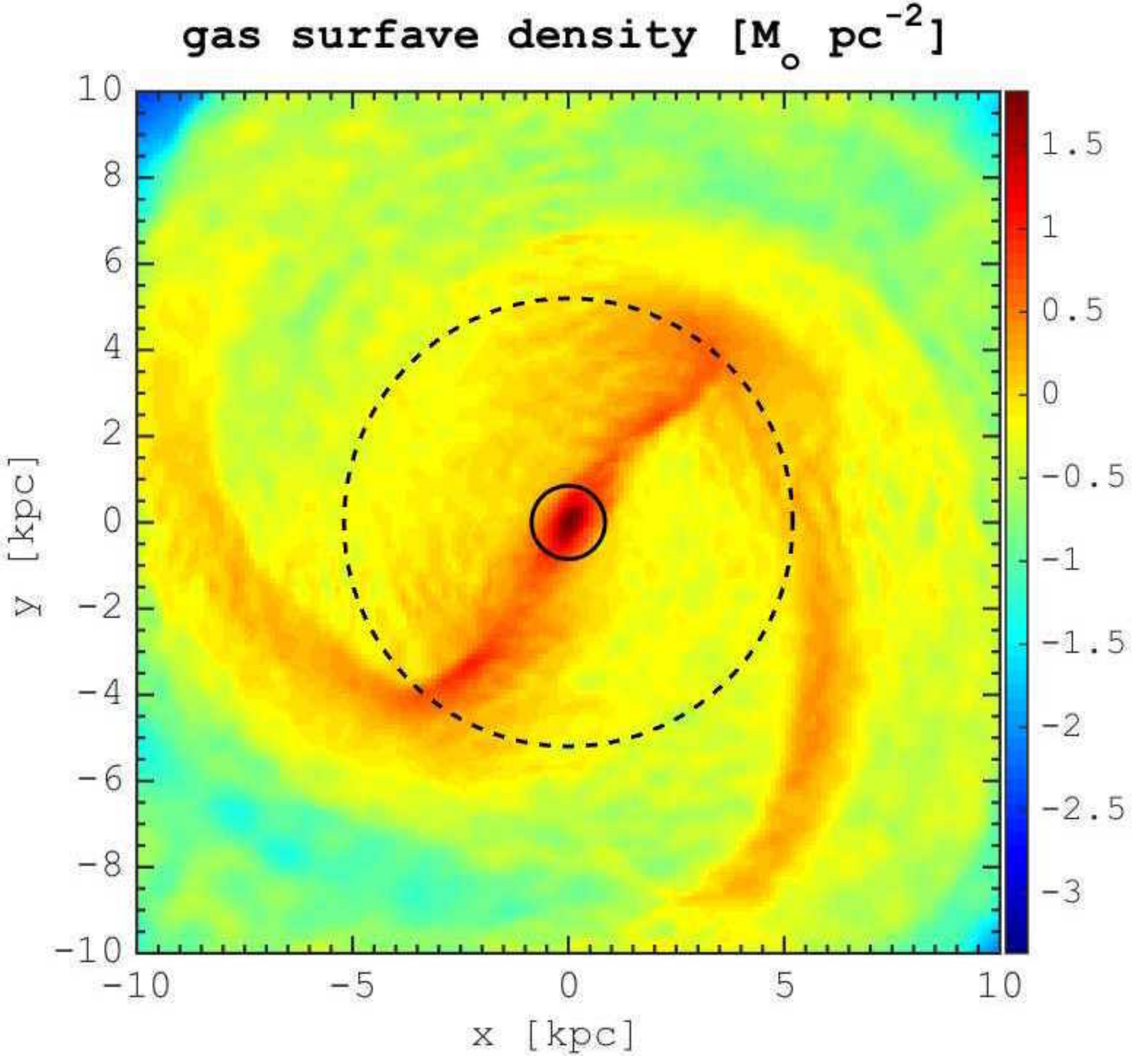}
\includegraphics[width=0.33\hsize]{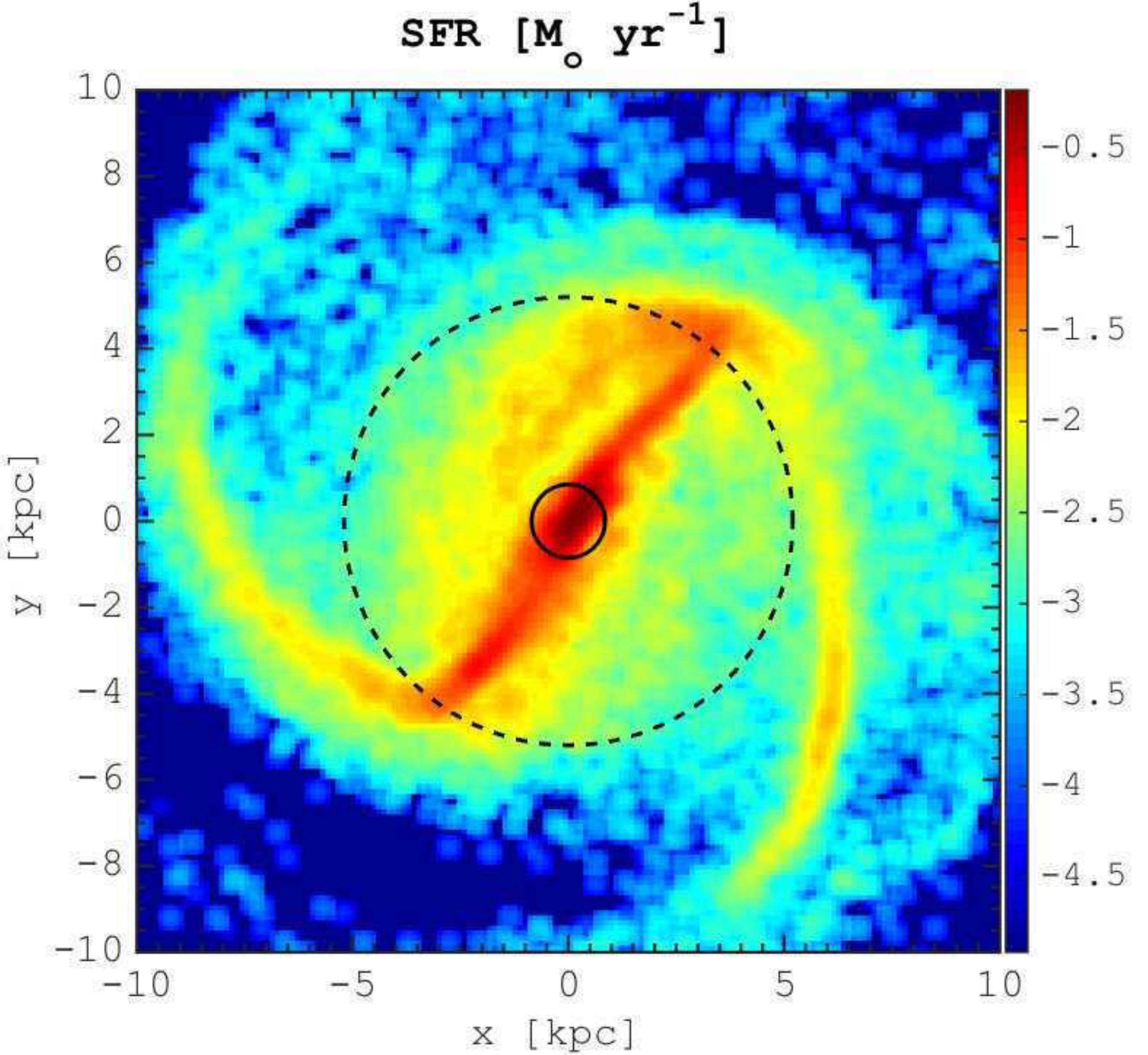}
\includegraphics[width=0.33\hsize]{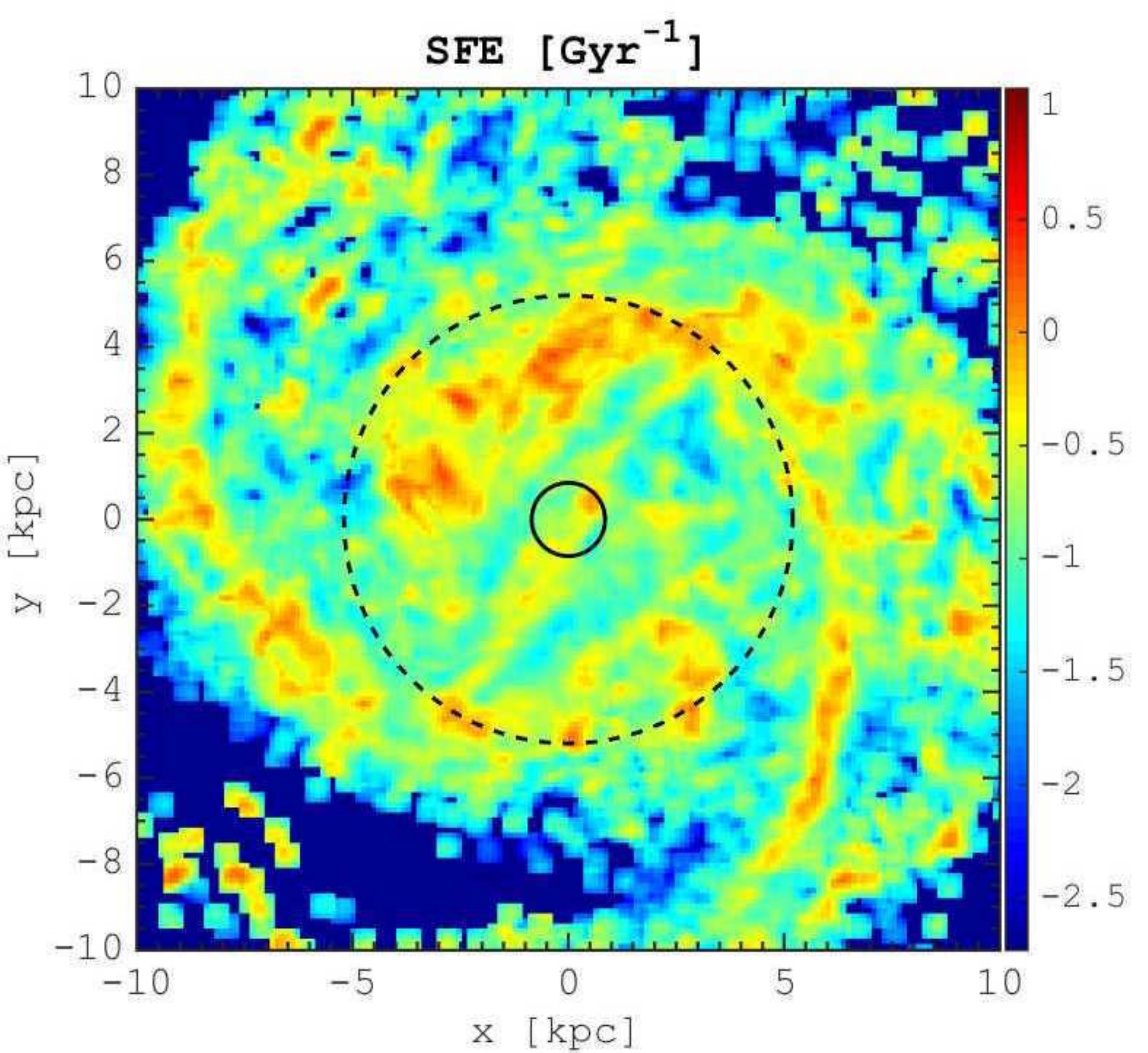}
\caption{Maps of the gas surface density~(left), surface star formation rate~(center) and star formation efficiency~(right) for barred galaxy model at a single time, $1.2$~Gyr. Black circles correspond to the positions of inner Lindblad resonance~(solid line) and corotation radius~(dashed line) for the bar.}\label{fig::sfe_map}
\end{figure*}
%%%%%%%%%%%%%%%%%%%%%%%%%%%%%%%%%%%%%%%%%%%%%%%%%%%%%%

Detailed analysis of the spatially resolved star formation efficiency in barred galaxy simulation is presented in Fig.~\ref{fig::sfe_map}. The locally averaged SFE in spiral arms is higher by a factor of $\approx 5$ than those of the entire disk region. The difference in gas dynamics between the bar and spiral arms may be the cause of this higher efficiency. The gas density and SFR are the highest in the nuclear region. SFE varies strongly with local conditions, but the locally averaged SFE in the nucleus is relatively low. The local values of SFE are constant in the circumnuclear overdensity and in the bar, and it increases toward the spiral arms. The comparison of SFE in the bar and spiral arms shows that SFE is about twice as high in the arms as those in the bar. Extreme SFE is found in the spiral arms, but not in the bar, indicates that the efficient triggering of star formation is related not only to the local gas density, but also to the local gas dynamics.

%Finally we conclude that star formation is more active in spiral arms and reduced significantly in bar. SFR map clearly shows this tendency, being about $5$ higher than the disk averages. The presence of the active star forming regions along the spiral arms confirms the visual impression that \citep{2017arXiv170606119G} found a remarkable agreement of the SFH of Sbc galaxies in CALIFA survey to SFH proposed for the MW. This comparison suggests that the star formation quenching can be a common phase early in the life of MW-like barred galaxies. 
%\cite{2015MNRAS.450.1375H} found the molecular gas depletion time in barred regions spans a range of more than a factor of $10$. 

\subsection{Dependence on bar strength and bar formation timescale}\label{sec::app1}

To generalize our results from the previous section, it is important to
understand how varying the bar parameters influence the suppression of
star formation in our models. To do this, we ran simulations with a range
of bar strengths $\varepsilon_b$ and bar growth timescales $h$. Importantly,
to investigate the influence of gas flows on the star-formation rate
in our simulations, we also ran simulations where we turned off the
converging flow criteria in the star-formation prescription. We note
here that the study of these additional models with different parameters
basically confirms the general trends we have found so far.

%%%%%%%%%%%%%%%%%%%%%%%%%%%%%%%%%%%%%%%%%%%%%%%%%%%%%%%%%%%%%%

\begin{figure*}
\includegraphics[width=1\hsize]{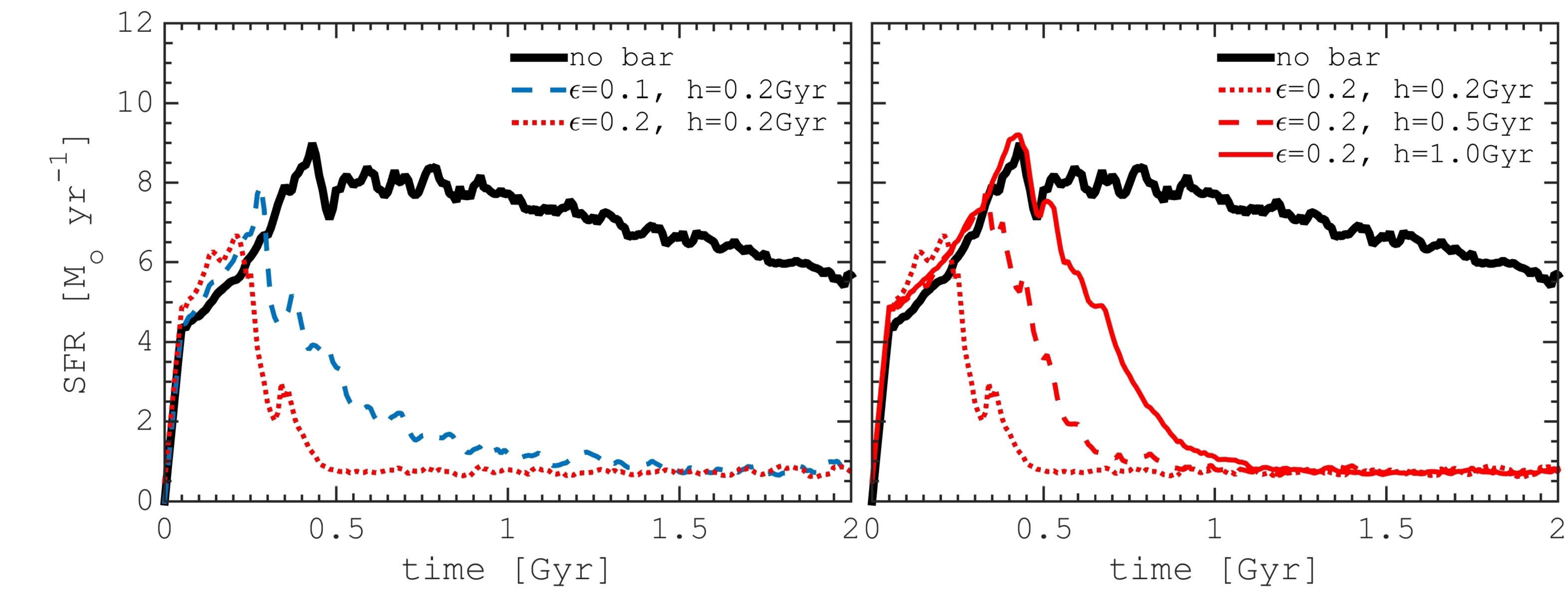}
\caption{Star-formation history in the different models. Unbarred galaxy
SFR is shown by black line in all panels. \textit{Left:} Comparison of
models with two different bar strengths, $\varepsilon_b = 0.1$ and $0.2$;
\textit{right:} Models with three different bar formation time-scales, $h = 0.2$, $0.5$,
$1$~Gyr. }\label{fig::app1_sfr_multi_models}

\end{figure*}

%%%%%%%%%%%%%%%%%%%%%%%%%%%%%%%%%%%%%%%%%%%%%%%%%%%%%%%%%%%%%%

Figure~\ref{fig::app1_sfr_multi_models} shows the SFHs in a variety
of disk simulations (see Table~\ref{tab::table_ini} for details).
In the model with the larger bar amplitude~($\varepsilon=0.2$), the bar
suppresses star formation earlier and even more rapidly. The influence
of the bar formation timescale is also very clear. The impact of the bar
occurs at later times for slower bar formation regimes and the timescale
of quenching itself is proportional to the bar formation timescale. In
all cases of the bar parameters we simulated\footnote{in models with the
standard SF prescription, see Fig.~\ref{fig::app1_sfr_multi_models}a,b},
the quenching of the SF  appears after about $1$~Gyr from the beginning
of the simulation, and coincides with the epoch of maximal growth of
the bar. After $1$~Gyr, all simulations show roughly the same SFRs,
equal to $\approx 1$~\Msunyr. This is a factor of $8-10$ lower than the
SFR of the unbarred model at the same time.

\subsection{Self-consistent simulations}\label{sec::app2}

\begin{figure*}
\includegraphics[width=0.5\hsize]{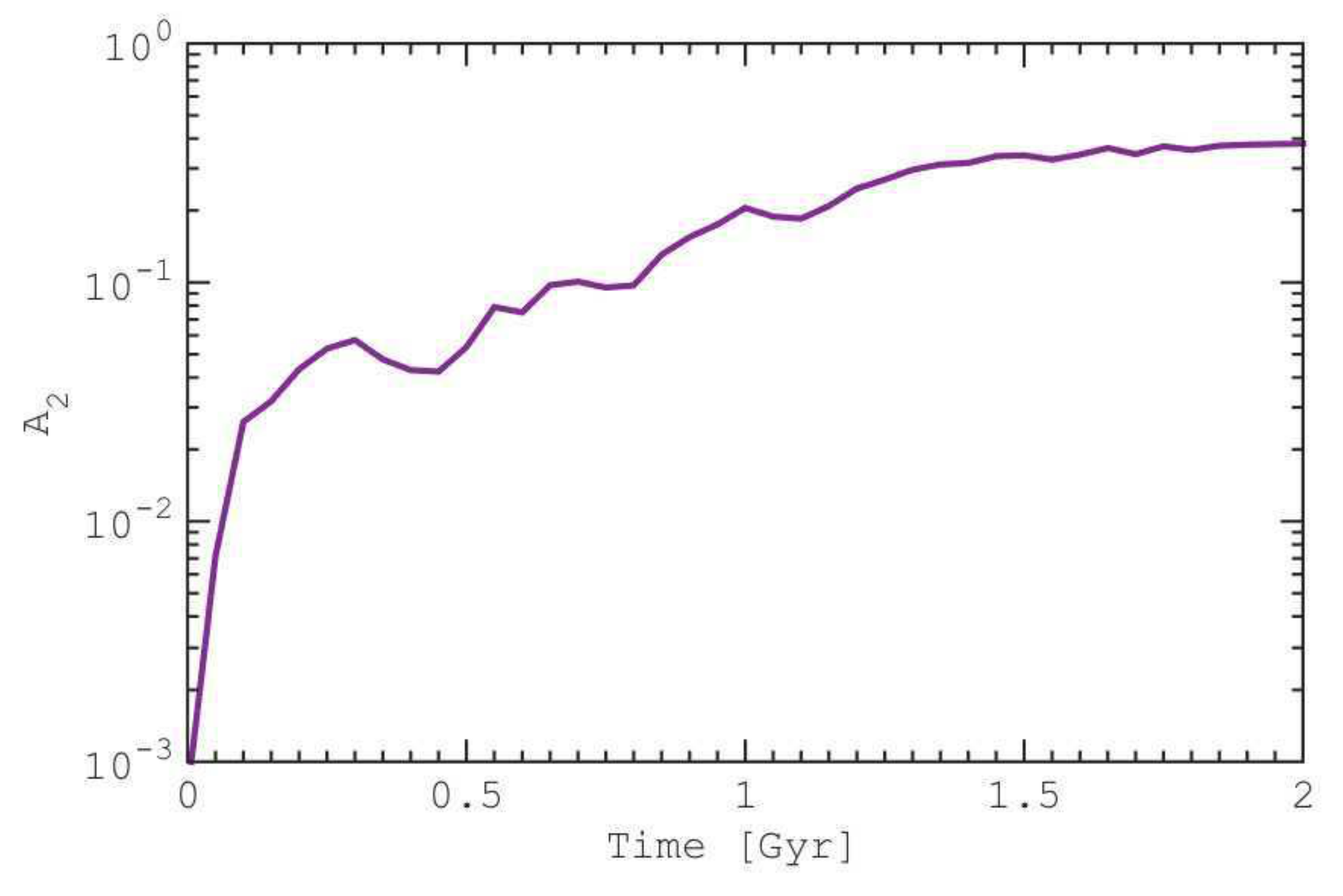}\includegraphics[width=0.5\hsize]{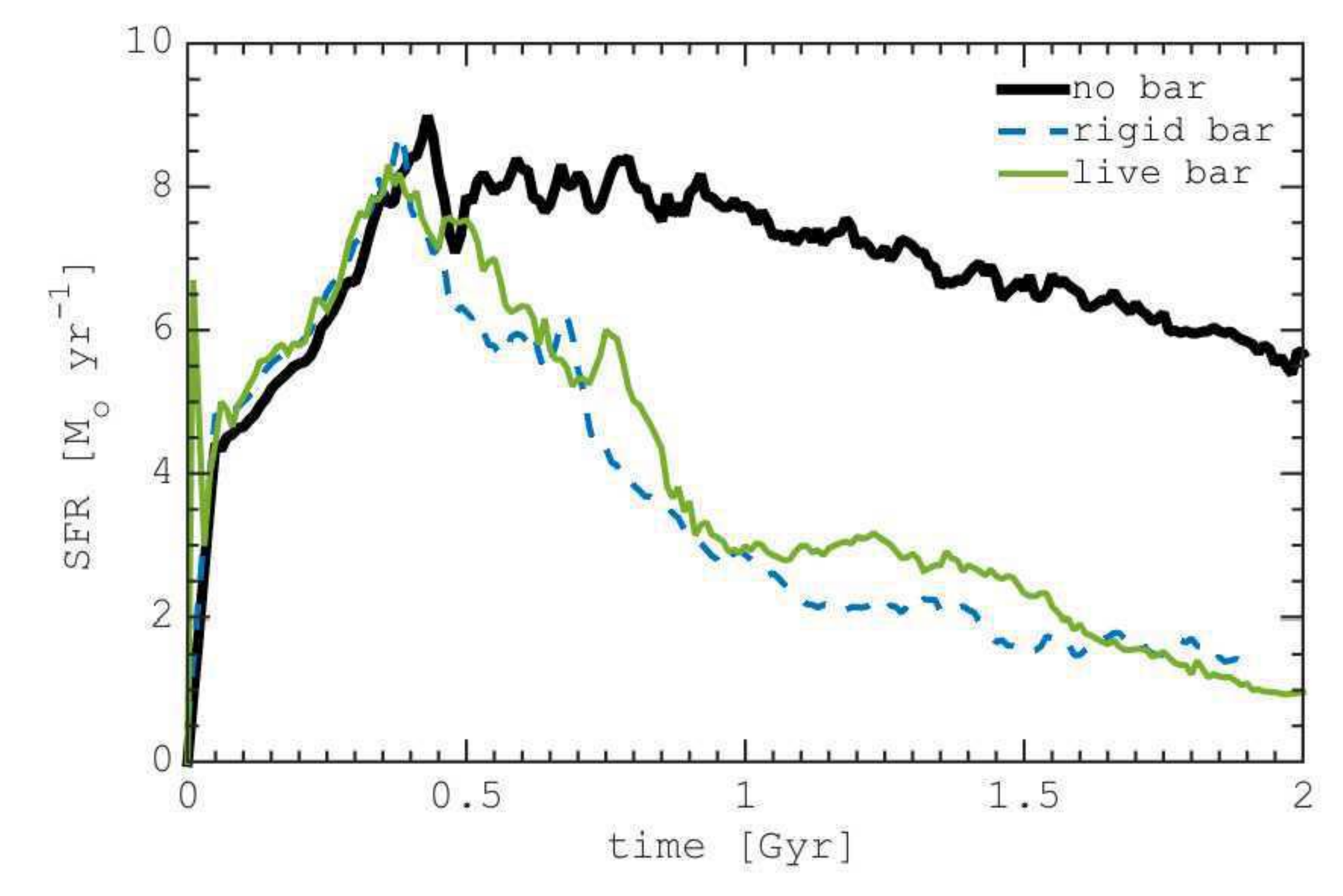}
\caption{\textit{Left:} Bar strength defined as the amplitude of the $m =
2$ Fourier moment, $A_2$, as a function of time in the self-consistent
model. \textit{Right:} Star-formation history in the live bar
simulation~(green solid line) and in the rigid bar simulation (dashed
blue line, model RBe02). The black line represents the star-formation
rate in model with no bar.}\label{fig::selfconsistent_models}
\end{figure*}

To further generalize our results, we also ran ``self-consistent''
simulations.  By self-consistent, we mean that the galaxy is composed
of a live stellar and gaseous disks, and a live dark matter halo. In
contrast to the simulations we have discussed, the bar potential is not
imposed analytically. In these simulations, a stellar bar naturally forms
after $0.5$~Gyr.  It arises as the result of energy and angular momentum
redistribution within the galaxy.  The disadvantage over our previously
discussed simulation is that we now need to estimate the bar strength,
which will generally depend on time. To quantify the strength of the
bar, we use the $m=2$ Fourier moment $A_2$ of the density distribution
in the following way:

\begin{equation}
A_2 = \sum^N_{k=1} m_k\exp{(2i\phi_k)}\,,\label{eq::A_2}
\end{equation}
where $m_k$ is the mass of $k$-th particle and summation is carried
out over both initial and recently formed stellar particles. We find
that the disc develops a strong bar whose length is $\approx 5$~kpc at
the end of simulation. The bar strength increases from the beginning
of the simulation and reaches a saturation level after about $1$~Gyr
(Fig.~\ref{fig::selfconsistent_models}).

Because of the smoother evolution of the bar amplitude,  the effect
of the bar on the gas dynamics is somewhat weaker in comparison to our
fiducial rigid bar model. However, it agrees very well with the model
RBe05~($h=0.5$~Gyr, $\varepsilon_b=0.2$) and in the self-consistent
isolated disk simulations we find the same trend in SFH as in the models
with a rigid potential (Fig.~\ref{fig::selfconsistent_models}). Our
self-consistent galaxy model confirms that star formation is suppressed
right after the bar strength reaches its saturation level. Star-formation
quenching is not as rapid as in our fiducial simulation with a rigid
potential, but it still appears after $0.5-1$~Gyr. The
gas velocity dispersion also increases and consequently star-formation
efficiency decreases. Indeed, there is a fast decrease of the SFR
from $8$ to $3$~\Msunyr during $\approx0.7$~Gyr, then the SFR is almost
constant for $0.5$~Gyr and then it decreases smoothly to $1$~\Msunyr.

%%%%%%%%%%%%%%%%%%%%%%%%%%%%%%%%%%%%%%%%%%%%%%%%%%%%%%%%%%%%%%

%%%%%%%%%%%%%%%%%%%%%%%%%%%%%%%%%%%%%%%%%%%%%%%%%%%%%%%%%%%%%%

\subsection{Star-formation prescription: the influence of the converging
flow criterion on star formation}\label{sec::sfr_prescription}

\begin{figure*}
\includegraphics[width=1.0\hsize]{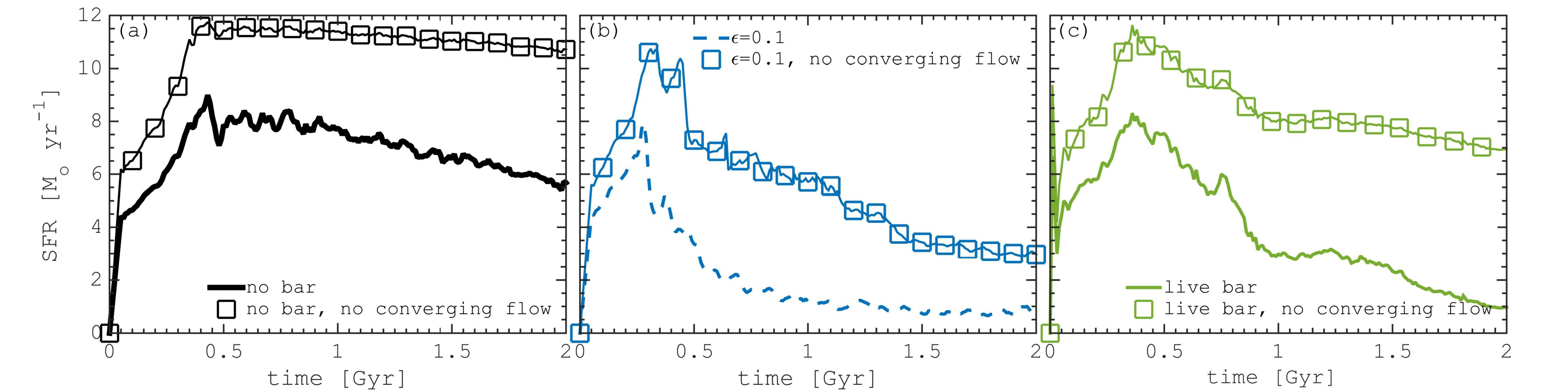}
\caption{Comparison of the star-formation rates in models with
and without converging flow criterion used in the star-formation
prescription. \textit{Left panel (a):} Models with no bar; \textit{center
panel (b):} models with a rigid bar ($\varepsilon=0.1$, $h=0.2$~Gyr);
\textit{right panel (c):} models with self-consistent model bar
formation.}\label{fig::conv_flow_impact}

\end{figure*}

Since our fiducial star-formation prescription includes the converging
flow criterion, the formation of stars directly depends on the gas
velocity dispersion. To investigate the influence of this particular
criterion, we have made three additional runs where we do not use the
converging flow criteria to form stars: model with no bar, rigid bar model
($\varepsilon_b=0.1$, $h=0.2$~Gyr), and self-consistent model with a live bar
formation (section~\ref{sec::app2}). Figure~\ref{fig::conv_flow_impact}
shows the evolution of star formation in these three models in comparison
to simulations where our fiducial star-formation criteria is used.

%%%%%%%%%%%%%%%%%%%%%%%%%%%%%%%%%%%%%%%%%%%%%%%%%%%%%%%%%%%%%%
%%%%%%%%%%%%%%%%%%%%%%%%%%%%%%%%%%%%%%%%%%%%%%%%%%%%%%%%%%%%%%

The models with the convergence flow star-formation prescription
demonstrate a very similar gas velocity dispersion evolution, but in
models with no converging flow criterion the star-formation rate is higher
by a factor of $2-6$ all the time. This behaviour was to be expected
because the conditions for the conversion of gas density into new stellar
particles are now weaker. For all models, star formation reaches up to
$11-12$~\Msunyr in  $\approx 0.5$~Gyr. In the unbarred galaxy model,
star formation slowly decreases only due to the conversion of gas into
stars, with a slower decay than in our basic no-bar simulation.

In barred galaxy simulations, star formation decreases not as rapidly
after the bar formation in comparison to our standard simulations. The SFR
decreases because the gas is driven to central regions and converted to
stars. In the standard model star formation is suppressed very rapidly
right after the bar formation while in a model where star formation
does not depend on the gas random motions star formation continues
after the bar formation, but at lower level decreasing only due to gas
consumption and depletion. Globally, in the case of no converging flow,
bar formation reduces the star-formation rate by a factor of few which is
much slower than in our fiducial runs. At the end of the simulation
(after $2$~Gyr), SFR is still high: $\approx 3$~\Msunyr for rigid bar
simulation and $\approx 7$~\Msunyr for self-consistent run. Even if a slow and weak decrease in the star formation rate is found also in the models  where the converging flow criterion is not implemented,  we conclude that the inclusion of random gas motions in star-formation prescriptions produces a rapid quenching phase rather than a slow star-formation rate decrease in models without such a prescription.

\subsection{Quenching parameters}\label{sec::quenching_params}

In this section we aim to quantify the star-formation quenching efficiency
in our various models and to investigate how the efficiency depends on
bar parameters, i.e., strength $\varepsilon$, timescale $h$. We introduce
simple quenching parameters. First, we estimate the quenching timescale
$h_q$ as an exponential timescale for the star-formation rate decrease:

\begin{equation}
{\rm SFR}(t) \propto \exp(- t / h_q)\label{eq::h_q}\,.
\end{equation} 

To make the fit, we did not use the whole time span of the simulations, but
only the period when star formation is suppressed. The second parameter
is the quenching rate $\zeta$ which we take as the ratio between the
maximum star formation rate before star formation suppression and its
value after the quenching phase:

\begin{equation}
\zeta = {\rm SFR(before\, quenching) / SFR(after\, quenching)}\,.\label{eq::quenching_rate}
\end{equation} 

\begin{figure*}
\includegraphics[width=1.0\hsize]{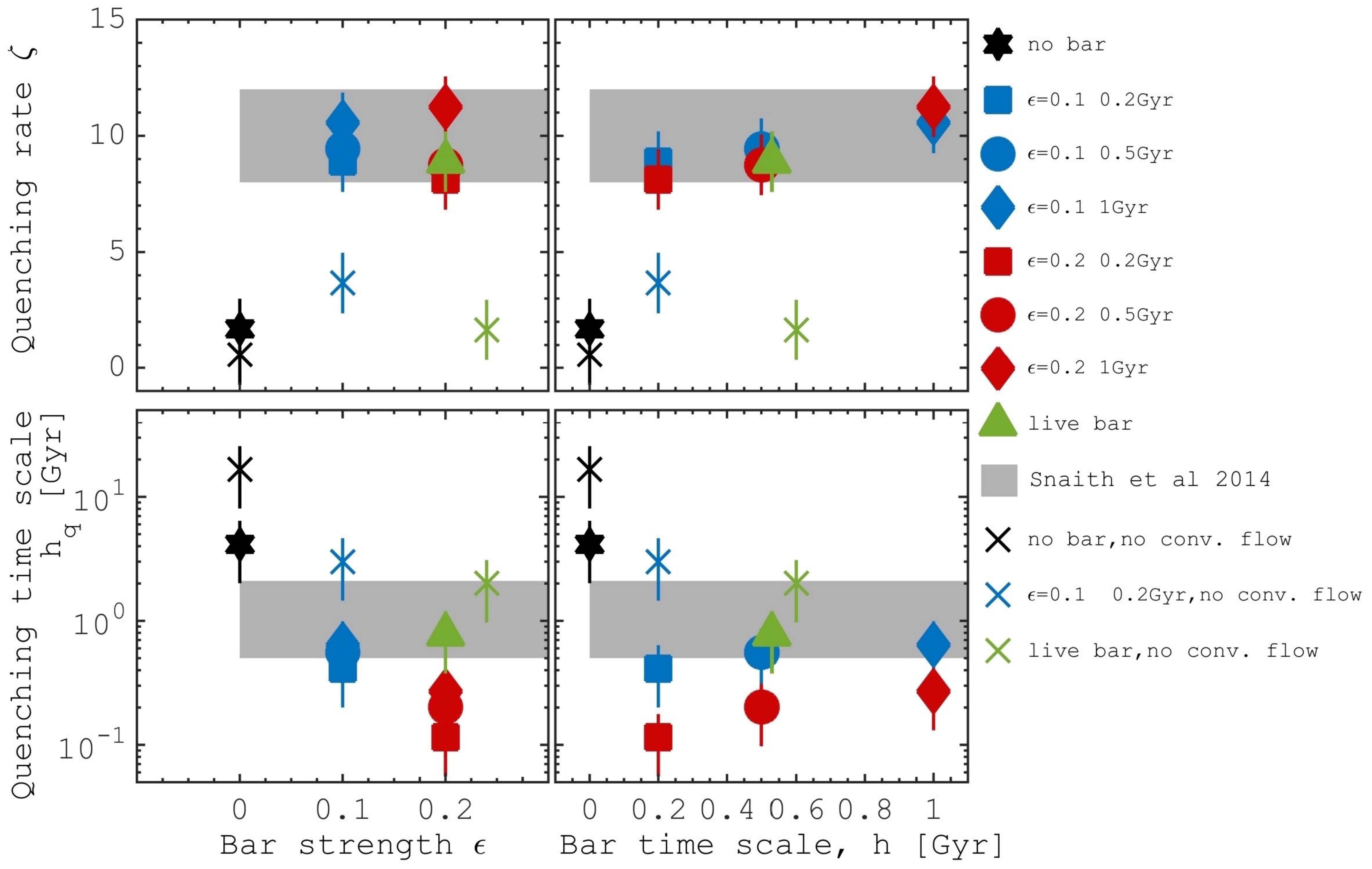}
\caption{Relation between the quenching rate, $\zeta$, and quenching
timescale, $h_q$, and bar parameters, bar strength $\varepsilon$ and bar
timescale $h$, in our simulated galaxies. Black 6-pointed star formally
shows the parameters measured for a model without a bar. The blue symbols
show the results for models with a bar strength $\varepsilon=0.1$,
while the red symbols indicate models with $\varepsilon=0.2$ with a
rigid bar potential. Green triangle corresponds to the results of the
self-consistent bar simulation. The filled grey area in each panel
represents the quenching parameters measured for star-formation
history of the inner Milky Way \citep{2014ApJ...781L..31S}. Note
that the symbol colors are the same as in star-formation
histories used in Figs.~\ref{fig::app1_sfr_multi_models} and
\ref{fig::selfconsistent_models}. Models without taking into account
converging flow criteria in the star-formation prescription are shows
by crosses. Error bars refer to the $95\%$~confidence interval from the
fitting procedure. }\label{fig::quenching_parameters}

\end{figure*}

For each model star-formation history we measure these two
parameters and analyze them as a function of the main bar parameters
(Fig.~\ref{fig::quenching_parameters}) For this analysis we also added
two models with no bar in order to distinguish quantitatively whether
quenching has occurred in a given run. Finally we estimate the quenching
parameters according to Eqs.~\ref{eq::h_q} and ~\ref{eq::quenching_rate},
for the Milky Way star-formation history derived from fitting the
solar vicinity age-$\rm [Si/Fe]$ abundance with a chemical evolution
model \citep{2015A&A...578A..87S}. Quenching parameters for the MW have
been estimated by analyzing the mean star-formation rates during the
thick~($9-11$~Gyr ago) and thin~($<8$~Gyr) disks formation phases taking
into account the estimated uncertainties in the SFR.  For the quenching
time scale, we find $0.6-2$~Gyr and for quenching rate, $8-12$.

%%%%%%%%%%%%%%%%%%%%%%%%%%%%%%%%%%%%%%%%%%%%%%%%%%%%%%%%%%%%%%

%%%%%%%%%%%%%%%%%%%%%%%%%%%%%%%%%%%%%%%%%%%%%%%%%%%%%%%%%%%%%%

We find that for all our standard star-formation prescriptions, barred
galaxy models produce a quenching episode with an exponential timescale,
$h_q = 0.1-0.8$~Gyr, and rate, $\zeta = 8-11$. For the unbarred simulations
these parameters are $4$~Gyr and $1.5$ respectively. For the runs
without converging flow criteria in the star-formation prescription,
the suppression of the star formation is much longer, $>2$~Gyr.

The process described here fits very well the observations of the
Milky Way: a decrease in the SFR by a factor of about $6-12$, occurring
in less than $1.5$ Gyr. Barred galaxy models with no converging flow
in star-formation prescription exhibits a much slower and less efficient star
formation decrease. Thus without taking into account the converging flow
criterion, we are not able to get a star-formation quenching comparable
to the Milky Way, quenching rate, $\approx 10$, and quenching timescale,
$\approx 1$~Gyr.

We clearly find that the stronger the bar, the faster star
formation is quenched (Fig.~\ref{fig::quenching_parameters}).
We find shallow linear dependences, $h_q \propto h$ and
$\zeta \propto h$ for our set of simulations with different
bar timescales and fixed bar strengths $\varepsilon = 0.1$ and
$0.2$ (Fig.~\ref{fig::quenching_parameters}). Interestingly, the
quenching rate does not depend on the bar strength, $\varepsilon$
(Fig.~\ref{fig::quenching_parameters}). It is possible that
during the bar amplitude growth, the star formation also increases
slightly, but quenching episode starts earlier for stronger bars
(Fig.~\ref{fig::app1_sfr_multi_models}).

\section{Discussion: Implications for the MW and distant disk galaxies}\label{sec::discuss}

The star-formation history of the Milky Way shows a quenching episode
between $10$~and $7$~Gyr \citep{2014ApJ...781L..31S, 2015A&A...578A..87S}.
Although this result was initially derived from stars in the solar
vicinity, this SFH has a more general validity as the stars originate
from all over the disk and yet, despite this ``geographical''
diversity, stars in the MW form a tight age-$\rm[\alpha/Fe]$
relation \citep{2014ApJ...781L..31S, 2015A&A...578A..87S}.
\cite{2016A&A...589A..66H} confirmed these results by showing that the
bimodality of the  $\rm[\alpha/Fe]$ distribution visible in the APOGEE
data in the entire inner disk can be reproduced if its star formation was
quenched at the end of the formation of the thick disk. This quenching
episode occurred throughout the inner disk ($\lesssim 7$~kpc of the galactic
center). In contradiction with many quenching models, the data suggest
however that the quenching did not occur because of the exhaustion of
gas. The chemical continuity observed on stars formed before and after
the quenching episode excludes the possibility that substantial amounts
of gas was accreted during this period.  This implies that the disk at
the end of the quenching epoch possessed a sufficient reservoir of gas
to resume star formation and form the thin disk.

\cite{2016A&A...589A..66H} suggested that the bar could be responsible
for the quenching over this epoch in the MW by increasing the
turbulence in the gas, preventing altogether the gas from collapsing,
becoming self-gravitating,
and forming stars. Indeed, the quenching of the star formation has
been observed in local galaxies despite them having substantial
gas reservoirs.  For example, \cite{2015MNRAS.448..258R} and
\cite{2015ApJ...801....1F} discovered significant molecular reservoirs
in post-starburst galaxies which is inconsistent with
their star-formation rate in comparison to normal galaxies. K-S
relation for some of these post-starburst galaxies suggests that the
star-formation efficiency is lower by a factor of $\approx20$ for a
given molecular gas mass.  \cite{2015ApJ...798...31A} proposed that
most of the molecular gas in NGC~1266 is very inefficient at forming
stars. The most likely explanation for the suppression of SF in dense
gas is that turbulence dissipation rate is long enough and the energy
high enough to keep the gas from collapsing gravitationally.  Thus the
sustained turbulence is a reasonable explanation for suppressing star
formation.  The cloud, $\rm G0.253+0.016$, is one example of where star formation is surpressed by turbulence.
This cloud is  located in the galactic center, has a lower star-formation
efficiency compared to similar clouds and has an especially high gas
velocity dispersion~\citep{2013ApJ...765L..35K}. These results and
others ~\citep[see, e.g.,][]{2015ApJ...812..117A, 2015A&A...574A..32G}
find that it is not necessary to substantially remove gas reservoirs to
quench or suppress star formation in galaxies on global or local scales.

We find that bars can be responsible for the increase of turbulence
in the gas, suppressing the star-formation efficiency within the
co-rotation radius. This scenario is different from the one proposed
by \cite{2015A&A...580A.116G}, where the star formation is quenched by
sweeping the gas with the bar to the Galactic center. As already discussed
in \cite{2016A&A...589A..66H}, we feel this solution is not appropriate
for the Milky Way, since once the gas is swept to the Galaxy center,
it needs to be replenished to allow the thin (very extended) disk to
form. The accretion of new gas would leave a signature in the abundances,
while the data shows strict continuity between the thick and thin
disks. On these grounds, such a model cannot be appropriate for the MW.

The possible role of bar in quenching processes was established empirically
by \cite{2010MNRAS.405..783M, 2012MNRAS.424.2180M}, who showed that for
a given amount of gas, barred galaxies are redder than unbarred galaxies. In
another words, disk galaxies with bars are more likely to be quenched than
unbarred galaxies \citep[see also][]{2011MNRAS.411.2026M}. Recently, also 
\cite{2016A&A...595A..67C} found that barred massive galaxies have
redder colors within their co-rotation radius. According to them this is consistent with bar-quenching, since bars bring gas to the center, and create nuclear concentration, and increase the bulge.

Because bars seem a very common phenomenon, occurring in two thirds
of disk galaxies, the process described here may be widely applicable.
Bar shocks injecting substantial turbulent energy into
galactic disk within bar co-rotation and the turbulence stabilize the
molecular gas against collapse. In this mechanism, the gas content from
galactic disks does not need to be expelled for galaxies to quench.
The underlying mechanism for bar quenching is simply to stir-up the gas
to high dispersions.  It is through these high velocity dispersions
that star-formation efficiency is decreased.  This mechanism,
high dispersions suppressing star formation has been confirmed both
empirically, phenomenologically, and numerically. However, many aspects
of bar quenching and the impact of turbulence on star formation remain
mysterious, such as understanding in detail how the bar rotational energy is
transferred to the gas on large and small scales, how rapidly this energy
dissipates as function of bar parameters, galaxy-type and mass, etc.

\section{Conclusions}\label{sec::concl}

This study emphasizes the fundamental role that stellar bars play in
influencing the gas dynamics on large scales within gas-rich galaxy
disks and thereby in modulating galaxy star-formation histories. Using
$N$-body/hydrodynamical simulations of gas-rich barred galaxies, we
find that star-formation history strongly depends on the formation and
evolution of bars. The results of our simulations show that it is possible
to suppress the star formation in barred galaxies by collecting most
of the gas in the bar-region and then by inducing significant random
gas motions through the shear and modulating gravitational potential
generated by the rotation of the bar. We find that the formation of
bars is responsible for quenching star formation in gas-rich disks of
galaxies. These findings strongly support the idea that a bar may have
had a substantial impact on the star-formation history of the Milky Way
\citep{2016A&A...589A..66H}.  \cite{2016A&A...589A..66H} found that a
decrease in the SFR by a factor of $10$ occurring within $1$ Gyr is needed
to explain the chemical abundances in the disk. They proposed that the
bar could be at the origin of the quenching by increasing the turbulence
in the ISM. At least with the detailed data we have on the Milky Way,
it appears that the rise of the stellar asymmetries in the disk could
be the origin of the transition from the thick disk to the thin disk
populations. The results above imply that it is essential to compute the
bar evolution and multi-phase ISM self-consistently in order to understand
star-formation history of spiral galaxies and their quenching phase.

From this study, we find that: 

\begin{itemize}

\item[$\bullet$] The formation of a bar plays a crucial
role in regulating the star-formation rates of massive gas-rich
galaxies. Our simulations demonstrate a rapid decreases of the
star-formation rate after the formation of the bar.

\item[$\bullet$] For our standard model of a bar with an amplitude of $10\%$,
the star-formation rate is suppressed by a factor of $10$ within $1$~Gyr
compared to a model with exactly the same parameters but has no bar.

\item[$\bullet$] We find that in barred galaxies gas velocity dispersion
strongly depends on the radius. For the central parts, meaning radii
within the length of the bar, the velocity dispersion can reach up to
$20-35$~\kmps. A large velocity dispersion is observed in the barred
galaxy simulations, even if there is no intense on-going star formation.

\item[$\bullet$] There is a significant growth of the gas velocity dispersion
within the co-rotation radius of the bar reduces the star-formation
efficiency by a factor of $2-5$.

\item[$\bullet$] We also demonstrated that the gas with high velocity dispersion
is maintained in the galactic disk. However, star formation is less
efficient in the turbulent ISM and in the majority of our models, the
star formation is quenched. Thus, we conclude that the action of the
bar can efficiently quench star formation without needing to deplete
the gas.

\item[$\bullet$] The analysis of the Kennicutt-Schmidt relation in simulated gas-rich galaxies reveals a flattening of its slope in the bar region. Currently observational studies on the spatially resolved star formation efficiency in gas-rich (or high-redshift) barred galaxies are missing, thus our conclusions can be used as a prediction to test with further observations.

\end{itemize}

\begin{acknowledgements}
We thank the referee for her/his helpful comments. The authors wish to thank N. Bastian, P. James, R. Schiavon, for enriching discussions. This work was granted access to the HPC resources of CINES under the allocations 2016-040507 (PI : F. Combes) and 2017-040507 (Pi : P. Di Matteo) made by GENCI.  This work has been supported by ANR (Agence Nationale de la Recherche) through the MOD4Gaia project (ANR-15-CE31-0007, P.I.: P. Di Matteo). Numerical simulations partially have been performed at the Research Computing Center~(Moscow State University) under the RFBR grant~(16-32-60043).  
\end{acknowledgements}

\bibliographystyle{aa} % style aa.bst
\bibliography{references} % your references Yourfile.bib

\end{document}